\ifx\compilefullpaper\undefined  
\documentclass[10pt, twocolumn, aps, nofootinbib, longbibliography, nobibnotes]{revtex4-1} 
\pdfoutput=1	
\usepackage{amsfonts, amssymb, amsmath, amsthm}
\usepackage{latexsym}
\usepackage[tracking=smallcaps]{microtype}	
\usepackage{url}
\usepackage{color}
\definecolor{DarkGray}{rgb}{0.1,0.1,0.5}
\definecolor{curgray}{rgb}{0.5,0.5,0.5}
\definecolor{curblue}{rgb}{0.04,0.11,0.64}
\definecolor{curpurple}{rgb}{0.65,0.16,0.58}
\definecolor{curorange}{rgb}{1,0.32,0}
\usepackage[colorlinks=true,breaklinks, linkcolor=DarkGray,citecolor=DarkGray,urlcolor=DarkGray]{hyperref}	

\usepackage{graphicx}
\usepackage{subfigure}

\newcommand{\ket}[1]{{|#1\rangle}}

\newcounter{sprows}

\newlength{\spheight}
\newlength{\spraise}

\newlength{\commentslength}

\newcommand{\rem}[1]{}

\newfont{\subsubsecfnt}{ptmri8t at 11pt}
\renewcommand{\subparagraph}[1]{\smallskip{\subsubsecfnt #1.}}

\newcommand{\eqnref}[1]{\hyperref[#1]{{(\ref*{#1})}}}
\newcommand{\thmref}[1]{\hyperref[#1]{{Theorem~\ref*{#1}}}}
\newcommand{\lemref}[1]{\hyperref[#1]{{Lemma~\ref*{#1}}}}
\newcommand{\corref}[1]{\hyperref[#1]{{Corollary~\ref*{#1}}}}
\newcommand{\defref}[1]{\hyperref[#1]{{Definition~\ref*{#1}}}}
\newcommand{\secref}[1]{\hyperref[#1]{{Sec.~\ref*{#1}}}}
\newcommand{\figref}[1]{\hyperref[#1]{{Fig.~\ref*{#1}}}}  
\newcommand{\figureref}[1]{\hyperref[#1]{{Figure~\ref*{#1}}}}  
\newcommand{\tabref}[1]{\hyperref[#1]{{Table~\ref*{#1}}}}
\newcommand{\remref}[1]{\hyperref[#1]{{Remark~\ref*{#1}}}}
\newcommand{\appref}[1]{\hyperref[#1]{{Appendix~\ref*{#1}}}}
\newcommand{\claimref}[1]{\hyperref[#1]{{Claim~\ref*{#1}}}}
\newcommand{\factref}[1]{\hyperref[#1]{{Fact~\ref*{#1}}}}
\newcommand{\propref}[1]{\hyperref[#1]{{Proposition~\ref*{#1}}}}
\newcommand{\exampleref}[1]{\hyperref[#1]{{Example~\ref*{#1}}}}
\newcommand{\conjref}[1]{\hyperref[#1]{{Conjecture~\ref*{#1}}}}

\allowdisplaybreaks[1]

\def\COLOR{}
\ifdefined\COLOR

\else

\fi

\definecolor{Cayenne}{rgb}{0.5,0,0}
\definecolor{Midnight}{rgb}{0,0,0.5}
\definecolor{Plum}{rgb}{0.5,0,0.5}
\definecolor{Teal}{rgb}{0,0.5,0.5}
\definecolor{Clover}{rgb}{0,0.5,0}
\definecolor{Maroon}{rgb}{0.5,0,0.25}
\definecolor{Ocean}{rgb}{0,0.25,0.5}
\definecolor{Tangerine}{rgb}{1,0.5,0}
\definecolor{Strawberry}{rgb}{1,0,0.5}
\definecolor{Fern}{rgb}{0.25,0.5,0}
\definecolor{Aqua}{rgb}{0,0.5,1}
\definecolor{Moss}{rgb}{0,0.5,0.25}
\definecolor{Mocha}{rgb}{0.5,0.25,0}
\definecolor{Lemon}{rgb}{1,1,0}
\definecolor{Asparagus}{rgb}{0.5,0.5,0}
\definecolor{Grape}{rgb}{0.5,0,1}
\definecolor{Iron}{rgb}{.3,.3,.3}
\definecolor{Steel}{rgb}{.4,.4,.4}
\definecolor{Purple}{rgb}{.5,0,.5}

\def\llbracket{{[\![}}
\def\rrbracket{{]\!]}}
\usepackage[utf8]{inputenc}
\usepackage[table]{xcolor}

\usepackage{enumitem} 
\usepackage{color}
\definecolor{cardinal}{rgb}{0.827, 0, 0}

\begin{document}

\fi

\vfuzz2pt 

\title{Distance-four quantum codes with combined postselection and error correction}
\author{Prithviraj Prabhu and Ben W. Reichardt}
\affiliation{University of Southern California}

\begin{abstract}

When storing encoded qubits, if single faults can be corrected and double faults postselected against, logical errors only occur due to at least three faults.  At current noise rates, having to restart when two errors are detected prevents very long-term storage, but this should not be an issue for low-depth computations.  We consider distance-four, efficient encodings of multiple qubits into a modified planar patch of the $16$-qubit surface code.  We simulate postselected error correction for up to $12000$ rounds of parallel stabilizer measurements, and subsequently estimate the cumulative probability of logical error for up to twelve encoded qubits.

Our results demonstrate a combination of low logical error rate and low physical overhead. For example, the distance-four surface code, using postselection, accumulates $25$ times less error than its distance-five counterpart.  For \textit{six} encoded qubits, a distance-four code using $25$ qubits protects as well as the distance-five surface code using $246$ qubits.

Hence distance-four codes, using postselection and in a planar geometry, are qubit-efficient candidates for fault-tolerant, moderate-depth computations.

\end{abstract}

\maketitle

\section{Introduction}
\noindent
\textbf{Error correction and postselection.}
Noisy Intermediate-Scale Quantum (NISQ) algorithms for eigensolvers~\cite{Peruzzo2014, Wecker15eigen} and machine learning~\cite{Biamonte2017} are growing as popular applications for state-of-the-art few-qubit quantum systems.  Unfortunately, these devices are still prone to large amounts of noise~\cite{mooney2021wholedevice, Google2019, Pogorelov21ionexp}.  Although error correction can decrease error rates~\cite{Shor95decoherence, Gottesman97thesis, terhal15review}, current experiments encode only one logical qubit that is still fairly noisy~\cite{honeywell21steane, Google2021, duke21BScode, Wootton20ignis}.
In this paper we simulate storing multiple logical qubits in a lattice, as a first step toward modeling few-qubit computations. We repeatedly correct and remove single-qubit errors.  On detecting a more dangerous---and less common---two-qubit error, we reject and restart.  This ``postselection" technique allows distance-four codes to achieve similar logical error rates to distance-five codes.  For example, as shown in \tabref{f:tab6lsumm}, a distance-four code can correct errors on six logical qubits with a similar failure rate as the distance-five surface code, using only $10 \%$ as many physical~qubits.  The table also shows that acceptance rates are fairly high, so occasional restarts should not be a major issue for low-depth NISQ~\mbox{algorithms}~\cite{cerezo2020variational, bharti2021noisy}.

\begin{table}
\label{f:tab6lsumm}
\vspace{-.3cm} 
\caption{Postselected error correction for $6$ logical qubits using $\llbracket 16,k,4\rrbracket$ codes on the $25$-qubit planar layout of~\figref{f:layout}. The probability of logical error, acceptance, and expected time to complete are shown for 300 time steps, with noise rate $p = 5 \times 10^{-4}$.  The $k = 6$ code achieves logical error rate close to the distance-$5$ surface code using only $10\%$ of the qubits.  In comparison, for $6$ physical qubits at memory error rate $p/10$, the probability of error is about~$6\times 300 \times p/10 = 0.09$.}
\hspace{-3.9mm}
{\includegraphics[width=.5\textwidth]{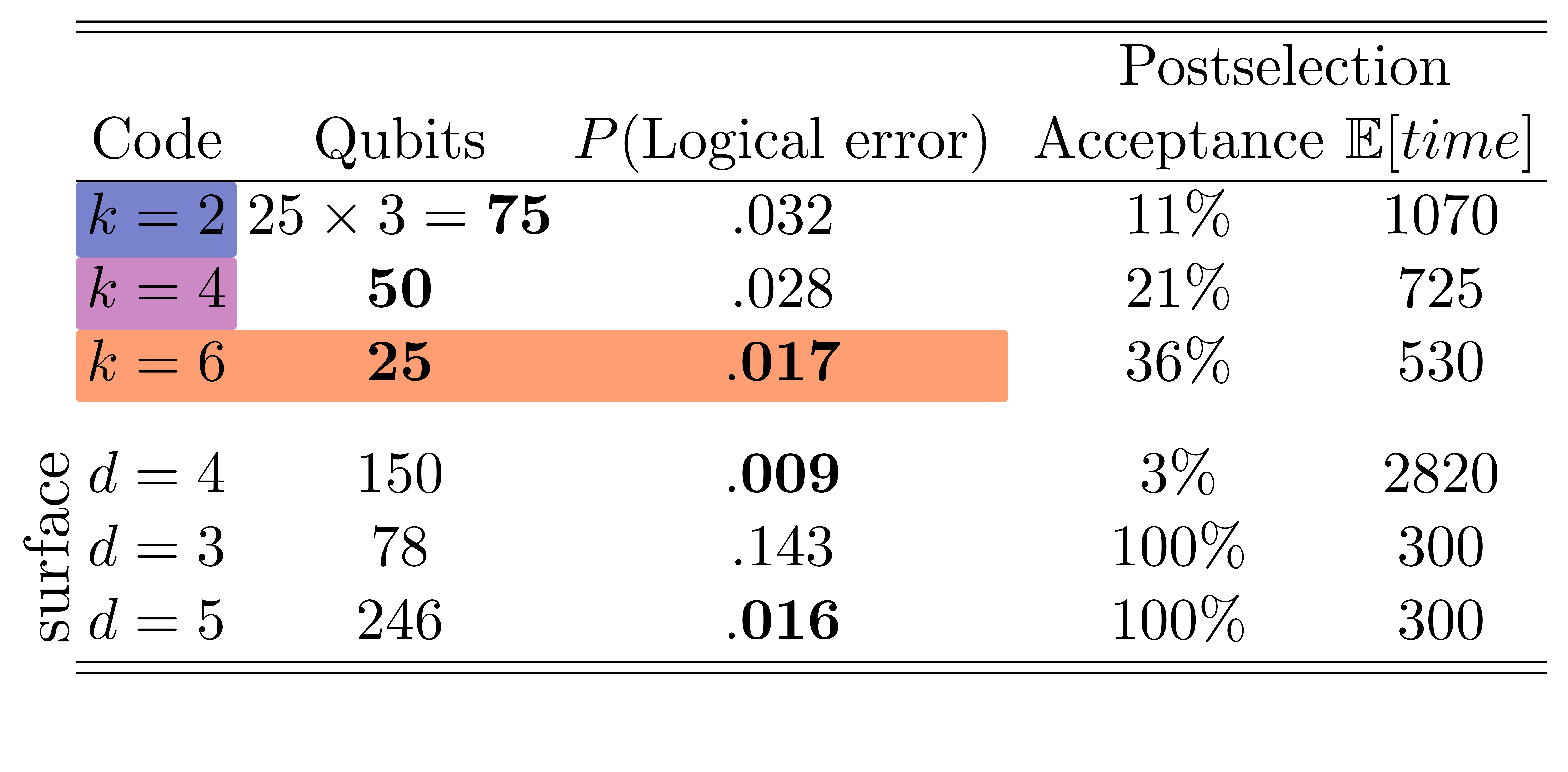}}
\vspace{-1cm} 
\end{table}

Postselection is a versatile tool in the quantum toolkit. 
In experiments, it has been used to decode the $\llbracket 4,2,2 \rrbracket$ error-detecting code~\cite{Linke17edexp, Takita17edexp} and the $\llbracket 4,1,2 \rrbracket$ surface code~\cite{Andersen2020, Google2021}.  In theoretical research, it has been used to reduce the logical error probability of state preparation~\cite{Paetznickgolay2013, ZhengBrunFTancprep18} and magic state distillation~\cite{ BravyiKitaev04magic, Meier13MSD}. Recently, postselection has been used to improve quantum key distribution~\cite{Yumang20QKDpost, Sekga21QKD} and learning quantum states~\cite{Scott18shadow, McClean2020}.  

\begin{figure}[b]
{\includegraphics[width=.38\textwidth]{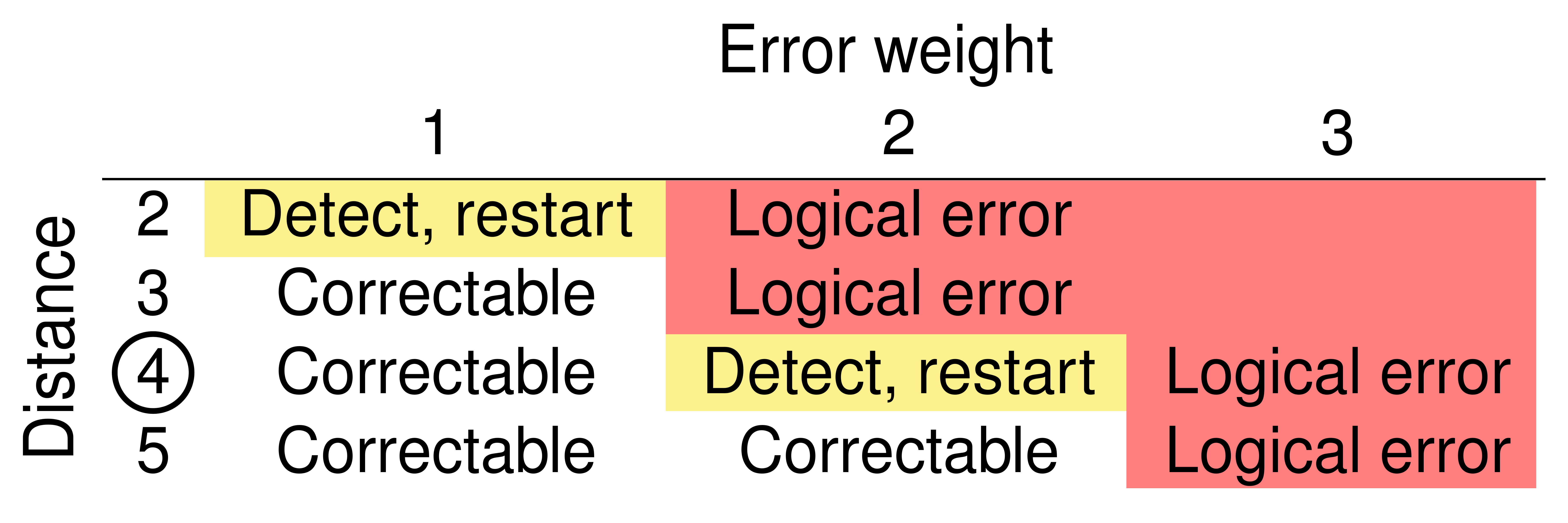}}
\caption{Distance-four codes with postselection lead to $O(p^3)$ logical errors, much like distance-five codes. 
Even-distance codes require restarts, however, unlike odd-distance codes.}
\label{f:disterrtab} 
\end{figure}

Knill previously combined postselection with error correction, on concatenated distance-two codes, to show an impressive $3 \%$ fault tolerance threshold~\cite{Knill05}.  
We also combine postselection and error correction, but with distance-four codes.
 As~\figref{f:disterrtab} indicates, distance-two codes can detect single errors and distance-three codes can correct them, meaning logical errors are due to second-order faults. Distance-three codes may alternatively be used to detect one or two errors, but then they lose the ability to correct and computations are very short-lived. We choose to use distance-four codes since they can simultaneously correct an error and detect two errors. Correcting some errors ensures restarts are less frequent, so longer computations can be run.  Since logical errors are caused only by third-order faults, logical error rates are very low.

\begin{figure*}
\includegraphics[width=.9\textwidth]{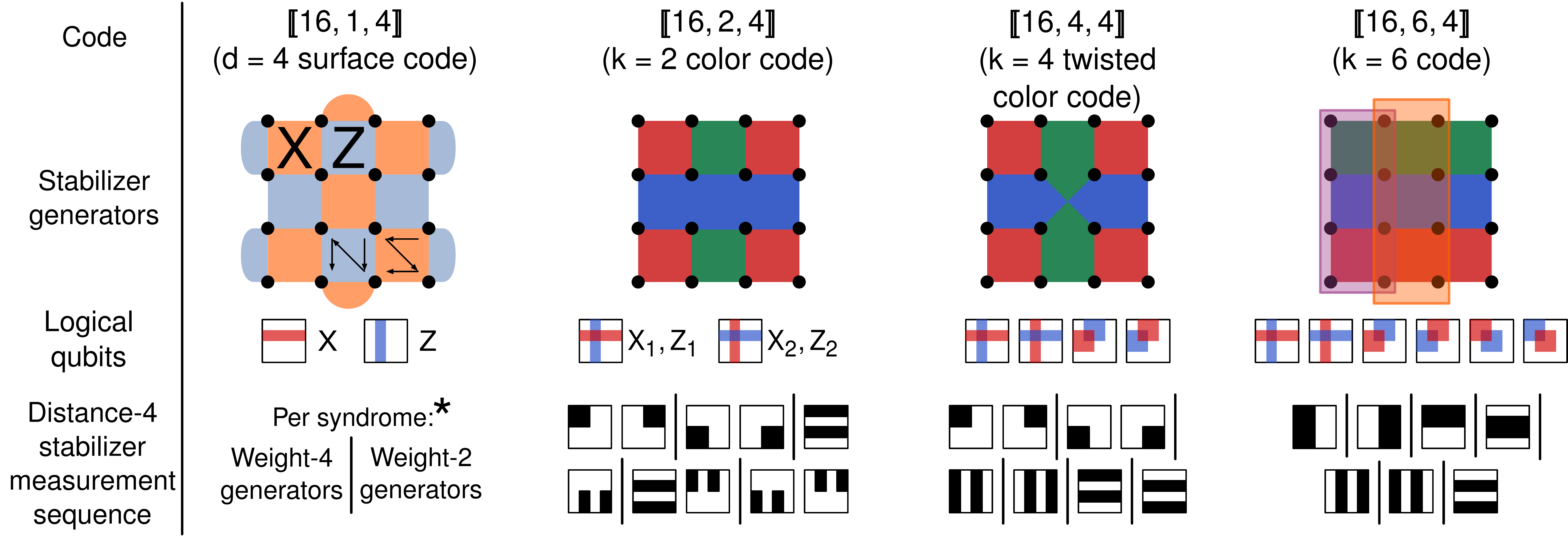}
\caption{Codes considered in this paper, with associated distance-four fault-tolerant $Z$ or $X$ stabilizer measurement sequences.  (The last three codes are self-dual CSS.)  Time steps of parallel measurements are separated by ``$\mid$".  ($\ast$)~For the surface code, fault-tolerant $X$ and $Z$ error correction is carried out using a rolling window of four syndromes, each measured in two time steps.}
\label{f:codes}
\end{figure*}

\begin{figure}
{\includegraphics[width=.14\textwidth]{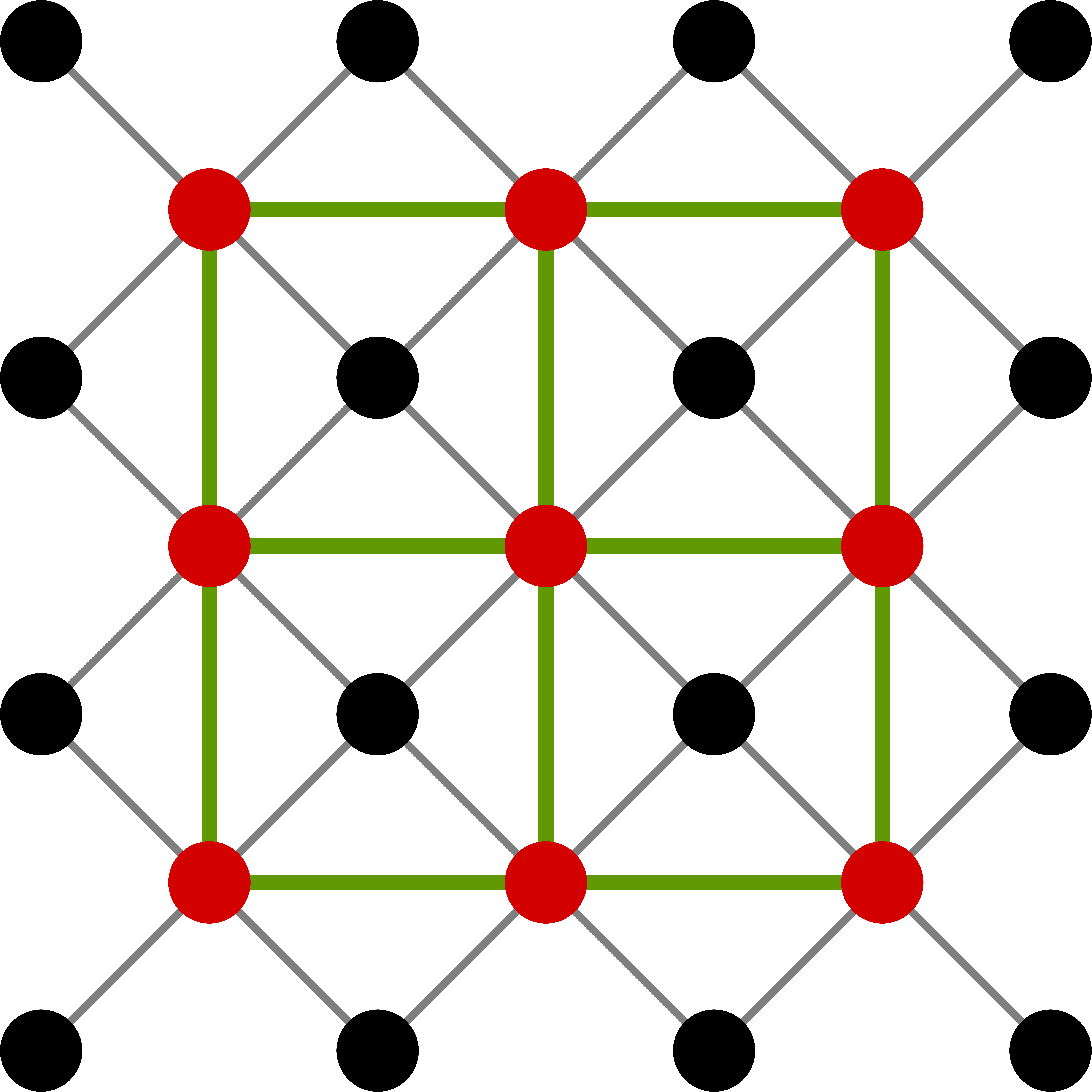}}
\caption{Planar layout of $16$ data and 9 ancilla qubits, in black and red respectively.  CNOT gates are allowed along the edges.  Grey edges are required for the surface code, and green edges between ancillas are required for the new codes in this paper.}
\label{f:layout}
\vspace{-0.3cm}
\end{figure}

\pagebreak 

\noindent
\textbf{Physical layout.}
In practice, it is difficult to build a quantum computer with native (fast, reliable) two-qubit gates between \textit{every} pair of qubits. Instead, qubits are placed on a one or two-dimensional lattice and two-qubit gates are mediated by local interactions, as in superconducting architectures and solid state systems. Current ion trap systems use long-range gates~\cite{Nigg14SteaneEC} and transport mechanisms~\cite{honeywell21steane} to connect all the qubits, but some degree of locality is required for larger systems.
In light of these connectivity constraints, it may be wiser to choose quantum codes that can be laid out on a lattice such that error correction requires the fewest number of local native gates.
The popular surface code has the attractive feature that it requires only nearest-neighbor interactions on a $2$D square lattice~\cite{Fowler12surface, CampbellroadstoFT2017}.  Similarly, error correction for topological codes has been investigated on sparser degree-three lattices~\cite{Chamberland20heavycodes, Chamberlandtriangularcodesflag2020, gidney2021faulttolerant}. 
But there is insufficient research on the performance gains of more densely connected layouts.  

We suggest using $16$-qubit codes on the $25$-qubit rotated square lattice of~\figref{f:layout}, where ancilla qubits additionally interact with neighboring ancillas.  This allows the use of 
flag qubits for fault tolerance~\cite{ChamberlandBeverland17flagft, ChaoReichardt17errorcorrection, chao2019flag, prabhu21}, in turn allowing measurement of large stabilizers. 
As shown in~\figref{f:codes}, we choose distance-four codes whose stabilizer generators are fairly local, with short Shor-style stabilizer measurement sequences that do not require any SWAP gates.
We consider two block codes and a color code that encode multiple qubits~\cite{delfosse2020short}, and the rotated surface code~\cite{BombinMartindelgado07surfaceoverhead} as a benchmark for postselection.  
In contrast, before the advent of topological codes, block codes were used for the simulation of $2$D local error correction~\cite{Svore07localSteane, Spedalieri09localBS, Lai14localKnill}.  These proposals performed Steane error correction on small distance-two and -three codes, and required many swaps.

\medskip
\noindent
\textbf{Results.} 
We compare our $16$-qubit codes with the $25$-qubit, distance-five surface code.
 We show below in~\figref{f:scaling} that, with rejection, the normalized logical error rate of the proposed codes is less than that of the distance-five surface code by as much as one order of magnitude. The distance-four surface code actually achieves two orders of magnitude separation. 

\begin{figure*}
\hspace{-5mm} 
\includegraphics[width =0.9\textwidth]{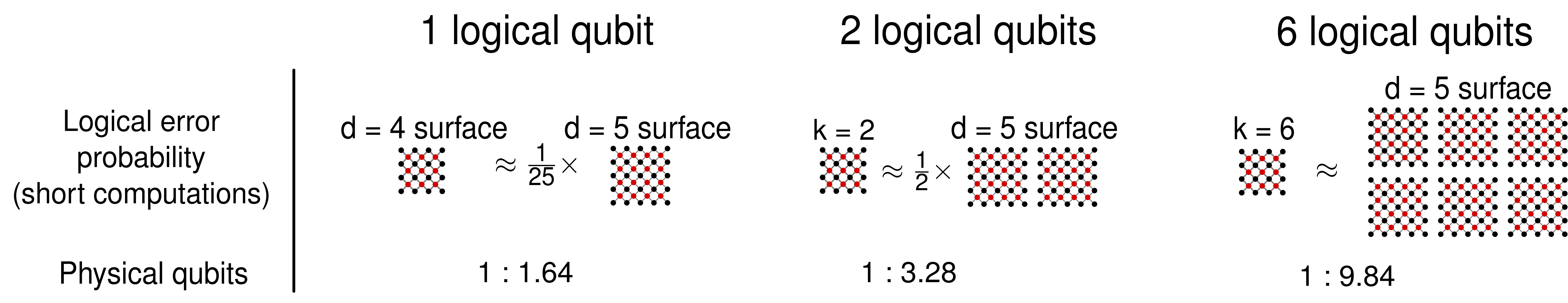}
\caption{Summary of results. For short computations, the probability of a logical error in the distance-$4$ rejection-based surface code is approximately $25$ times lower than that of the distance-$5$ variant. Further, for $6$ logical qubits, the $k=6$ code on one patch of $25$ qubits can match $6$ patches of the distance-$5$ surface code.}
\vspace{-0.1cm}
\label{f:resultssumm}
\end{figure*}

However, the logical error rate per time step does not capture the drawback of restarts. Instead, a better metric is the cumulative probability of logical error.
\figureref{f:resultssumm} compares this metric between the different codes for short computations that do not restart too often (more information in~\figref{f:PLEt}).
For one logical qubit, the distance-four surface code vastly outperforms its distance-five counterpart, and the $k = 2$ and $k = 4$ codes achieve a good balance of low qubit overhead and low logical error rate.
 We also show that just $50$--$75$ physical qubits are sufficient for good protection of twelve logical qubits.  Overall, we obtain lower logical error rates with higher encoding rates, using postselection and multi-qubit codes.

\medskip
\noindent
\textbf{Future work.} 
In order to verify these results on current quantum systems, some work is required.  Dense qubit connectivity in ion trap systems may allow for simple measurement of high-weight stabilizers, but superconducting devices generally prefer low qubit degree due to high crosstalk.  It may be possible to modify the circuits in this work to allow maximum qubit degree at most five or six, such as in the IBM Tokyo device~\cite{Tan21optimality}.
Consequently, in~\figref{f:degree4k2} below, we show that error correction of the $k=2$ code is possible with degree-four connectivity, but requires many extra qubits.

We only show how to do fault-tolerant error correction, but the ultimate goal is to perform
quantum computation. 
Selective logical measurements could induce computation within a patch, and transversal gates between vertically stacked code patches could facilitate non-Clifford gates.  
If these operations introduce a low amount of error, it may be possible to execute relatively high-depth circuits. These tools can then be used to execute short NISQ and magic state distillation algorithms.  As an example, our results show that just $50$ physical qubits may be sufficient to demonstrate $10$-to-$2$ MEK distillation experimentally with $O(p^3)$ logical errors~\cite{Meier13MSD}.  

\medskip
\noindent
\textbf{Organization.}
In \secref{s:codes} we provide more details about distance-four codes and the examples we choose in this paper. \secref{s:ftec} details the methods used for fault tolerance. In particular, stabilizer measurement circuits are dealt with in \secref{s:SMC} and sequences of stabilizer measurements are handled in \secref{s:SMS}. The noise model and results of simulations are contained in~\secref{s:results}. \secref{s:future} concludes with a discussion of future work and open questions.

\section{Codes} \label{s:codes}

We compare the error correction performance of six $\llbracket n,k,d \rrbracket$ stabilizer quantum codes, where $n$ is the number of physical data qubits and $k$ is the number of logical qubits.
A distance-$d$ quantum code should correct all errors of weight $j \leq t= \lfloor \frac{d-1}{2} \rfloor$, occurring at rate $O(p^j)$, for error rate $p$.
At low error rates, an unlikely error of weight-$(d-j)$ may be misidentified as the more likely weight-$j$ error, inducing a logical flip on recovery.
In even-distance codes, errors of weight $d/2$ 
can be detected, but applying a correction may induce a logical flip.
In this paper, we stop the computation instead of attempting to correct, ensuring logical flips only occur at rate $O(p^{t+2})$ and not $O(p^{t+1})$ as before.

For the distance-four codes shown in~\figref{f:codes}, we show that the logical error rate scales as $O(p^3)$ like a distance-five code.  
As a benchmark we first consider the rotated distance-four surface code~\cite{tomita14surface} on the layout of~\figref{f:layout}.  For a fair comparison of both the resource requirements and logical error rate, we consider additional benchmarks: the  distance-three and -five surface codes.  As in Ref.~\cite{tomita14surface}, each distance-$d$ surface code uses $d^2$ data qubits and $(d-1)^2$ ancilla qubits.  

The next three codes are the central focus of this work.  These self-dual CSS (Calderbank-Shor-Steane) codes were first considered in Ref.~\cite{delfosse2020short}, to show examples of codes that can be constructed to have single-shot sequences of stabilizer measurements. 
By fixing some of the logical operators, the $k=6$ code can be transformed into the $k=4$ and $k=2$ codes. Alternatively, puncturing the $k=6$ code yields the well-known $\llbracket 15,7,3 \rrbracket$ Hamming code.

\medskip
\noindent
\textbf{Improvements.}
Although these codes encode more logical qubits, they suffer from the difficult task of having to measure weight-eight stabilizers.  It is possible to construct a $\llbracket 16,2,4 \rrbracket$ subsystem code with only weight-four stabilizers and gauge operators.  Using the layout of~\figref{f:layout}, we compared  this code with the $k=2$ subspace code in this paper but found no significant improvements. This code is still useful, however, as we show in~\secref{s:future}.

Many other codes can also be constructed with $16$ qubits. For a biased-noise system, a CSS code with two logical qubits can be constructed with $Z$-distance six and $X$-distance four.  For more logical qubits, a non-CSS $\llbracket 16,7,4\rrbracket$ code can be used~\cite{Grassl07codetable}.
Although its stabilizer generators are larger, flag-based measurement may still offer a low-overhead route to fault-tolerance.

\section{Fault-tolerant error correction} \label{s:ftec}

A stabilizer measurement \textbf{circuit} is made fault tolerant to quantum errors by using extra physical qubits.  These ancillas are used to catch faults that may spread to high-weight errors.  In contrast, the bad faults in a syndrome extraction \textbf{sequence} flip syndrome bits.
Additional stabilizers are measured, essentially encoding the syndrome into a classical code.

A circuit is fault-tolerant to distance $d$ if $j \leq t= \lfloor \frac{d-1}{2} \rfloor $ mid-circuit faults cause an output error of weight at most~$j$.  
Additionally for even distance fault tolerance, sets of $d/2$ faults spreading to weight $>d/2$ errors should be detected so the computation can be restarted.  
When these faults yield an error of weight $d/2$, the computation is restarted if the faults can be detected, else it is rejected in the next round of error correction.

\subsection{Stabilizer measurement circuits} \label{s:SMC}

Quantum error correction involves the measurement of a set of operators called stabilizers, to diagnose the location of errors. For fault-tolerant error correction, these stabilizers may be measured individually, as in Shor's scheme~\cite{Shor96}, or together, using Steane- or Knill-type syndrome extraction~\cite{Steane97, Knill03erasure}.  

The flag method is a popular spin-off of Shor's scheme~\cite{ChamberlandBeverland17flagft, ChaoReichardt17errorcorrection, chao2019flag, prabhu21}.  By connecting multiple data qubits to each flag qubit, large stabilizers can be measured with relatively low overhead.  In addition, flag circuits can be made fault-tolerant only up to a desired degree. For example, Shor-style measurement of a weight-$w$ stabilizer needs $w+1$ ancillas and is fault-tolerant to distance $w$, but we show a weight-eight stabilizer measurement circuit with six ancillas that is fault-tolerant to distance four.

\begin{figure}
\subfigure[\label{f:smc00} ]{\includegraphics[width=.124\textwidth]{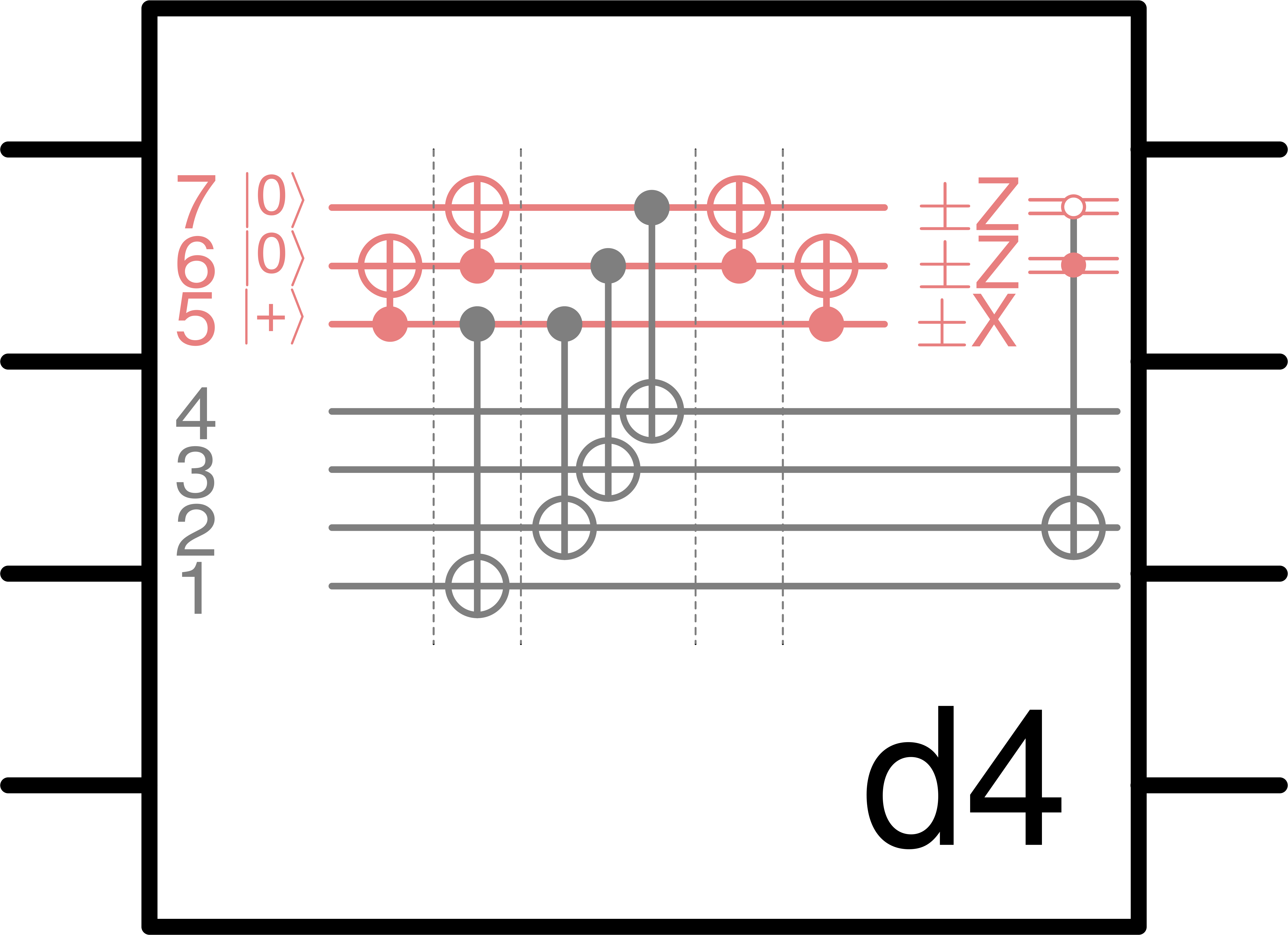}}
\hspace{5mm}
\subfigure[\label{f:smc0102} ]{\includegraphics[width=.315\textwidth]{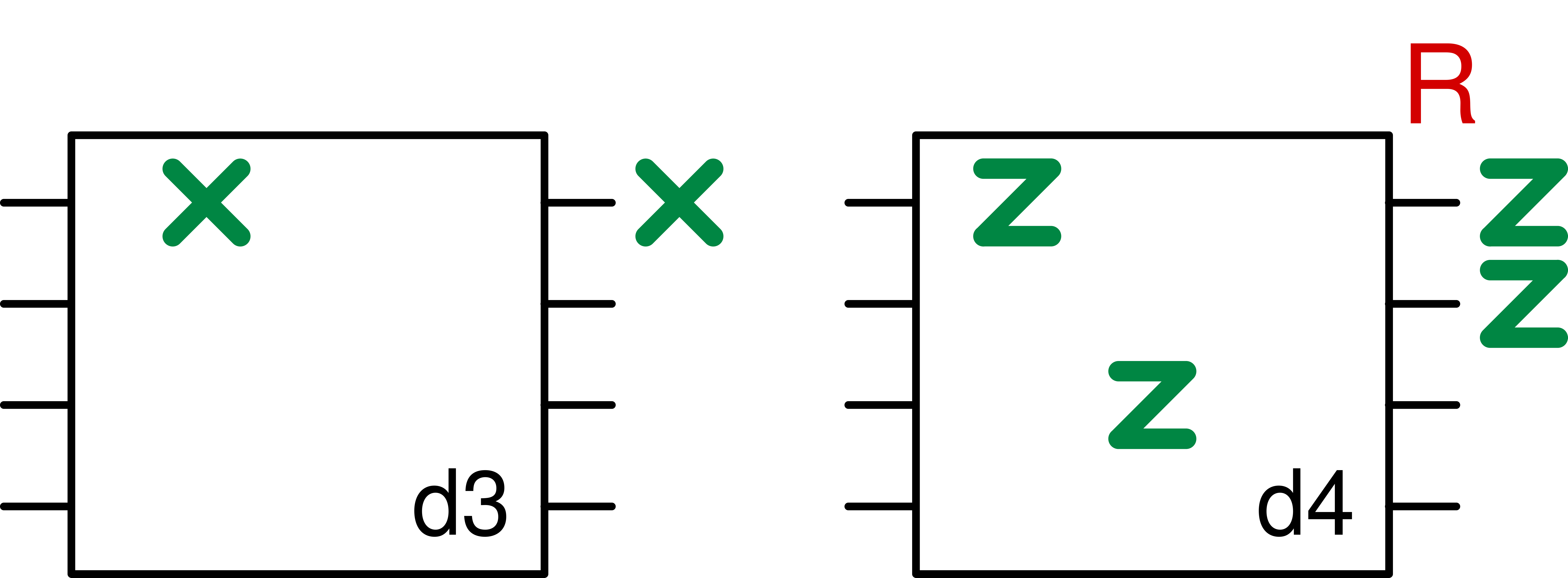}}
\caption{(a)~A distance-$4$ stabilizer measurement circuit contains ancilla preparation, CNOTs, measurement and a recovery.  (b)~Rules for fault tolerance.  One fault should be corrected to an error of $X/Z$ weight at most one---this is sufficient for distance $3$.  Two faults should either be rejected ({\color{cardinal} R}) or result in an error of weight~two.}
\label{f:smcrules}
\end{figure}

For distance-three fault tolerance, we show in~\figref{f:smcrules} that one fault in the circuit should result in an error of $X$ and $Z$ weight at most one.  For distance four, if two faults occur and can be detected, the computation must be rejected and restarted.  If this detection is not possible, the circuit must be designed to ensure errors cannot spread to weight greater than two.  Note that a fault may alter the value of the measured syndrome bit; syndrome bit errors are dealt with in~\secref{s:SMS}.

\begin{figure}
\hspace{-2.5mm}
\subfigure[\label{f:wt4circ} ]{\includegraphics[width=.23\textwidth]{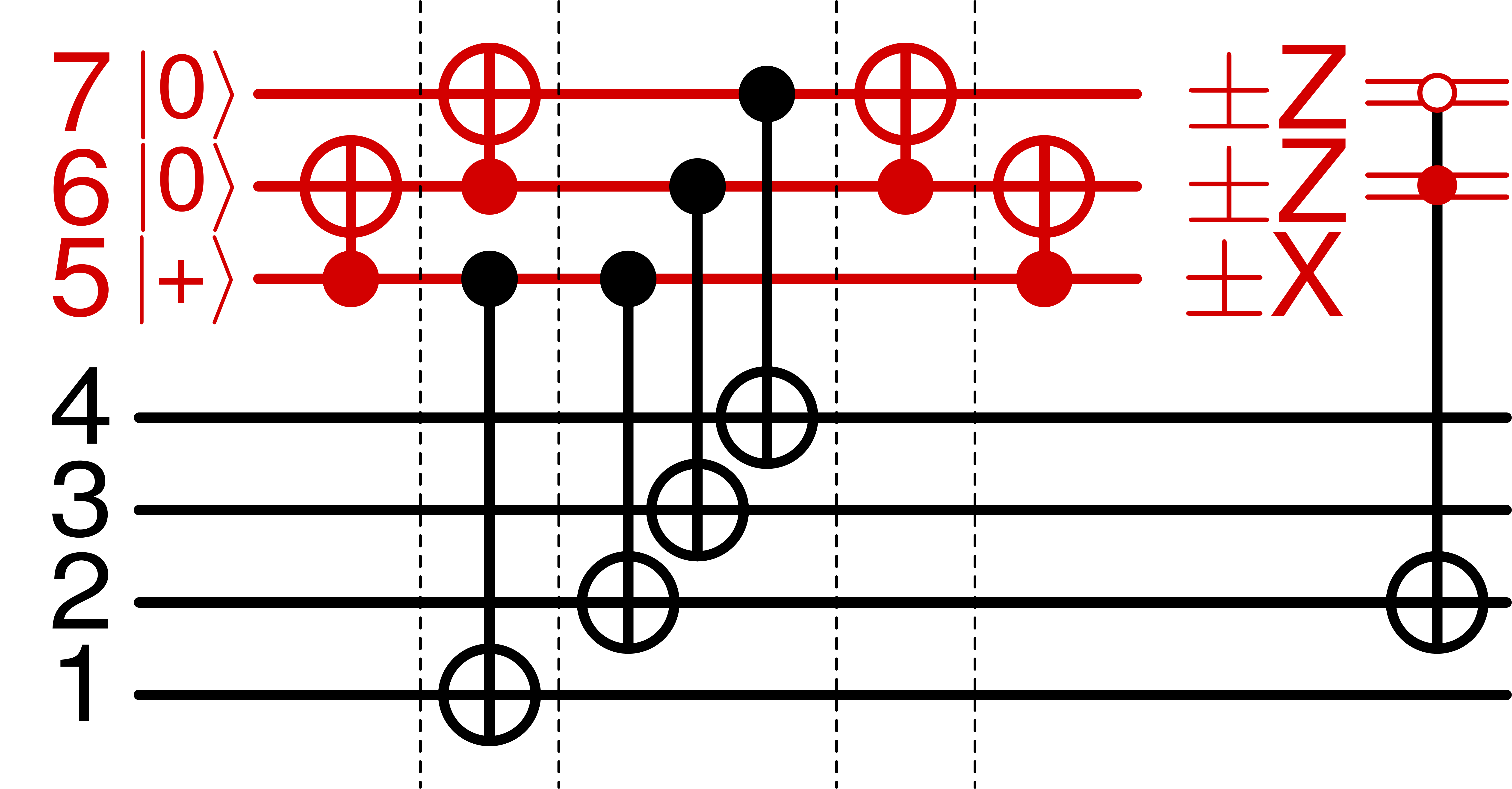}}
\hspace{1.5mm}
\subfigure[\label{f:wt4layout} ]{\includegraphics[width=.24\textwidth]{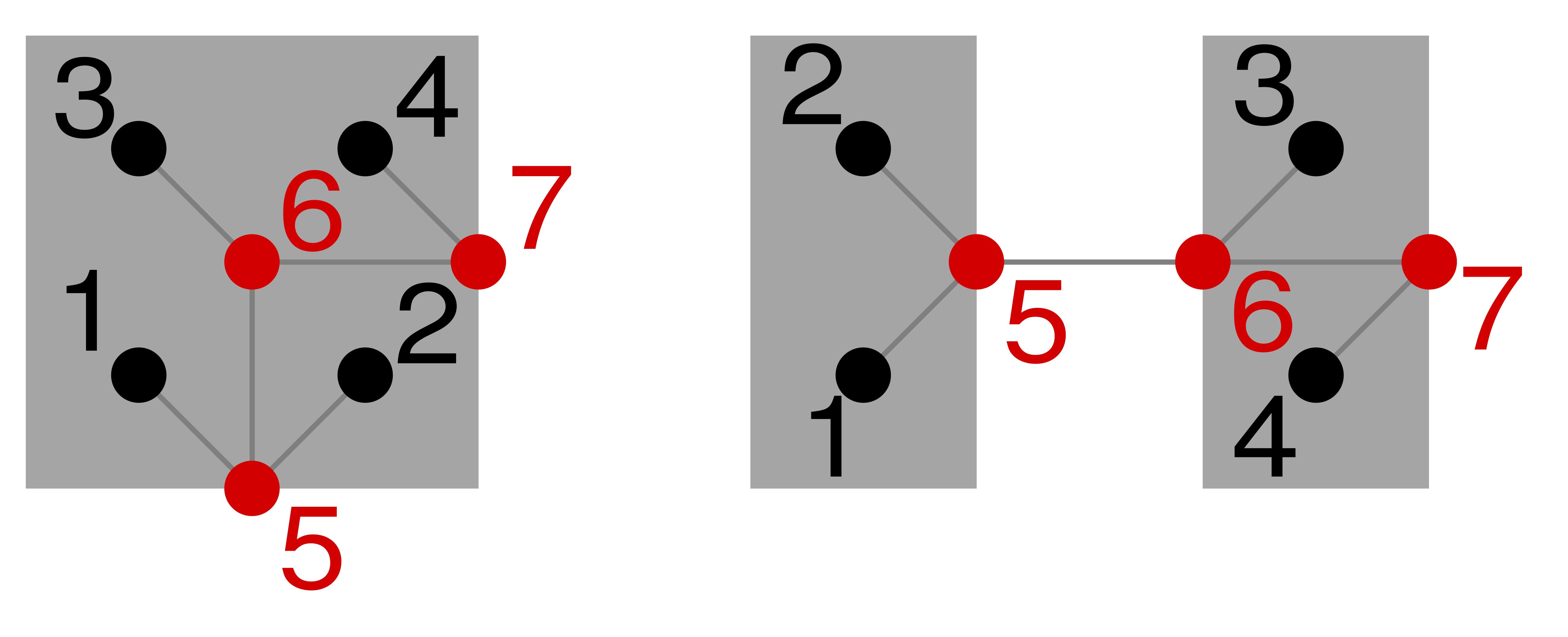}}
\caption{(a) Circuit to measure a weight-four $X$ stabilizer fault-tolerantly to distance-four, satisfying the locality constraints in~(b).  The $\pm Z$ measurements are used to flag mid-circuit faults. Gates bunched together can be performed in parallel. (b)~Two layouts for measuring stabilizers in the sequences of~\figref{f:codes}.}
\label{f:wt4circuit}
\end{figure}

\begin{figure}
\subfigure[\label{f:wt8circ} ]{\includegraphics[width=.44\textwidth]{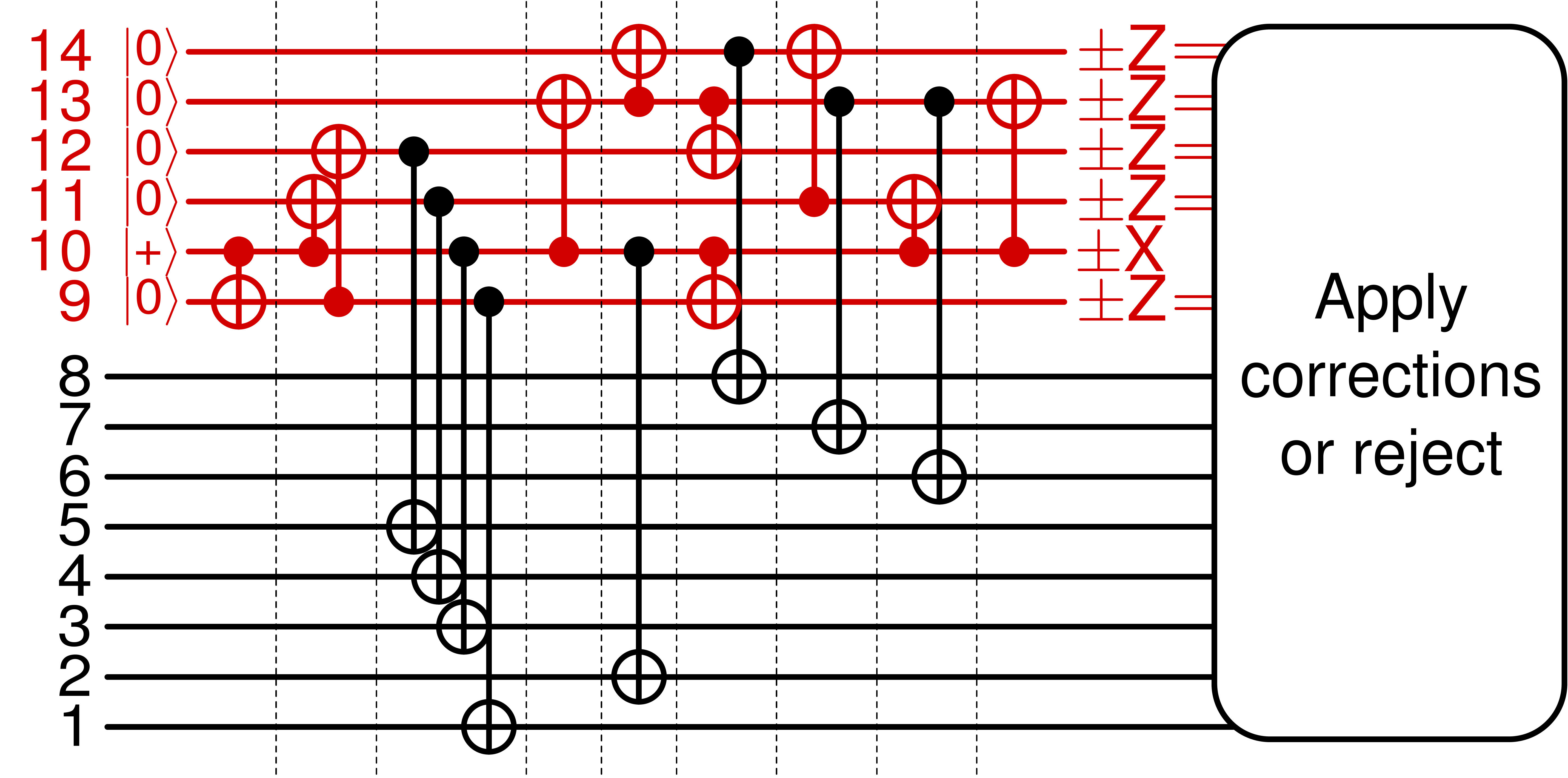}}
\subfigure[\label{f:wt8layouta} ]{\includegraphics[width=.43\textwidth]{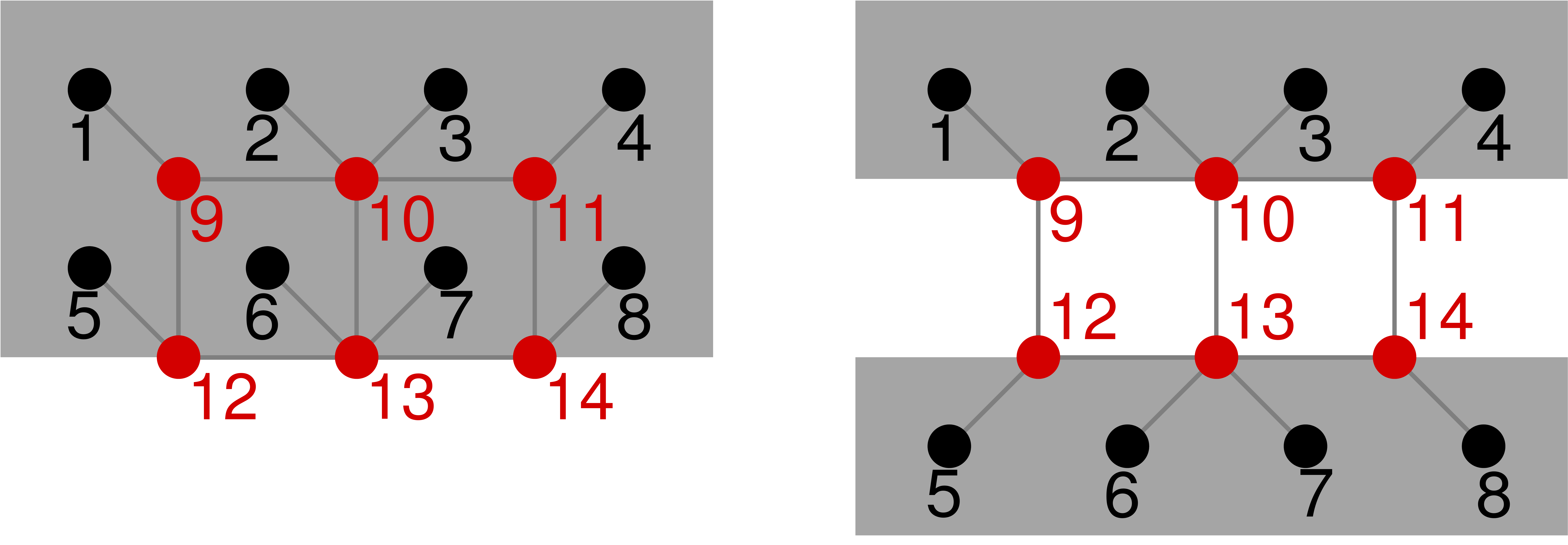}}
\caption{(a) Circuit to measure a weight-eight $X$ stabilizer fault-tolerantly to distance-four, satisfying the locality constraints in~(b).  One fault is corrected to at most a weight-one error, but two or more faults may either be corrected, or detected resulting in rejection.  The resulting flag outcomes for corrections and rejection are tabulated in \appref{s:wt8corrs}.  (b)~Two layouts for measuring stabilizers in the sequences of~\figref{f:codes}.}
\label{f:wt8circuit}
\end{figure}

We develop flag-based stabilizer measurement circuits.
For this, we use a randomized search algorithm constrained by the above fault tolerance rules and the geometric locality of~\figref{f:layout}.  With all six codes in this work, the stabilizers that are measured are of weight two, four and eight.  At the circuit level, the measurement of a weight-two stabilizer is automatically fault-tolerant (one fault causes an error of weight at most one).  
A weight-four stabilizer measured fault-tolerantly to distance three (i.e., one fault results in error of weight at most one) is automatically fault-tolerant to distance four, as two faults occurring in the circuit cannot create data errors with $X$ and $Z$ weight greater than two.  In~\figref{f:wt4circ}, the weight-four stabilizer measurement circuit applies a correction only for the $01$ ancilla measurement.
\figureref{f:wt8circ} shows a novel circuit to measure weight-eight stabilizers fault-tolerantly to distance four. This circuit uses different patterns of flag-qubit measurements to either correct an error, or reject---detecting an $O(p^2)$ fault event.  The flag patterns associated with corrections or with rejection have been tabulated in~\appref{s:wt8corrs}.  Figures~\ref{f:wt4layout} and~\ref{f:wt8layouta} show different ways of arranging the qubits to measure weight-four and weight-eight operators.

\medskip
\noindent
\textbf{Improvements.}
The benefit of measuring stabilizers individually is that error decoding is relatively simple. When stabilizers with overlapping support are measured in parallel, as in the surface code, more complicated decoding algorithms like minimum-weight perfect matching are required. However, we can still make small improvements for additional parallelism. In the $k=2$ and $k=4$ codes, only two of the corner weight-four stabilizers can be measured simultaneously, as each stabilizer requires three ancilla qubits. We conjecture that by sharing one ancilla qubit among all the corner stabilizers, it may be possible to fault-tolerantly measure all four of them using just nine ancilla qubits, like in Ref.~\cite{Reichardt18steane}.  Alternatively, in Steane-style syndrome extraction, subsets of stabilizers are measured in parallel using $n$-qubit resource states. Nine ancilla qubits are not sufficient for Steane's method, but Ref.~\cite{Huang21ShorSteane} shows that any subset of stabilizers can be jointly measured with specific resource states. If the fault-tolerant preparation of those resource states is possible on the nine-qubit ancilla sub-lattice of~\figref{f:layout}, it will be possible to develop faster and more efficient stabilizer measurement circuits.

The circuit of~\figref{f:wt8circ} uses six ancilla qubits for distance-four fault tolerance.  We found a distance-three fault-tolerant circuit using only four ancillas (qubits $9, 10,11,13$ in~\figref{f:wt8layouta}), which also requires fewer rounds of parallel gates. 
On the layout of~\figref{f:layout}, this may free up enough ancillas to measure two weight-eight stabilizers in each time step. The result is that more stabilizers can be measured faster and data qubits in an error correction block experience less idle noise.  Since these circuits are only fault-tolerant to distance three, a future avenue of research could use techniques in Ref.~\cite{ChaoReichardt17errorcorrection} to look at their performance in adaptive distance-four error correction. 

\vspace{-0.3cm} 
\subsection{Stabilizer measurement sequences}  \label{s:SMS}

\begin{figure}
\includegraphics[width =0.48\textwidth]{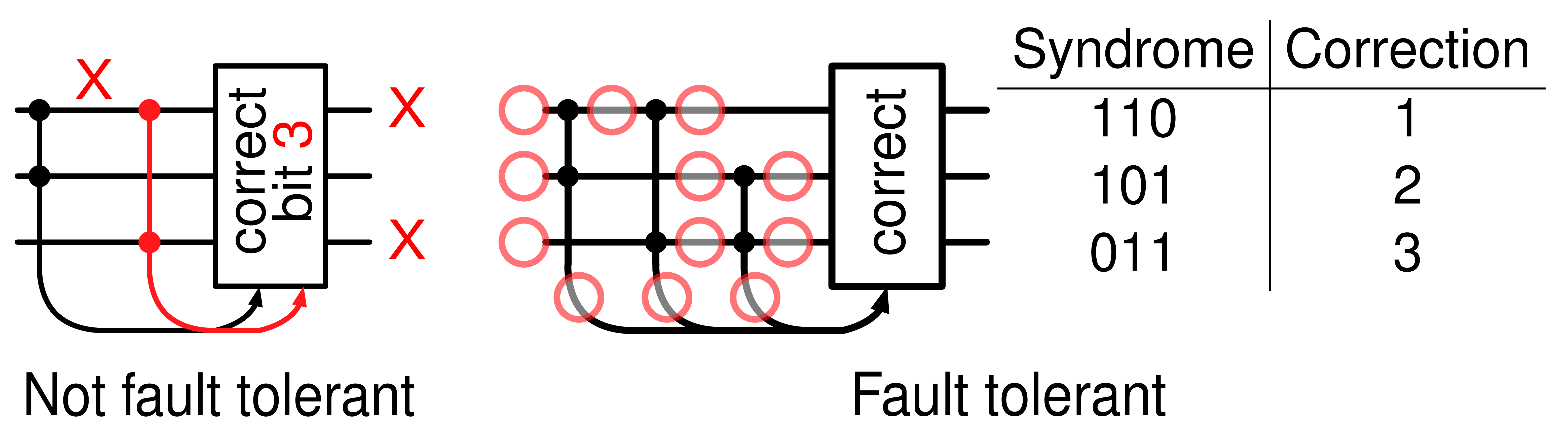}
\caption{Fault-tolerant error correction with the three bit repetition code $\{ 000, 111\}$ (adapted from~Fig. 1 of Ref.~\cite{delfosse2020short}).  It is not fault tolerant to correct errors based on the two parity measurements $1 \oplus 2$ and $1 \oplus 3$. An internal error on bit $1$ can be mistaken for an input error on bit $3$, as they yield the same syndrome.  Errors can be corrected fault-tolerantly by adding another parity check, $2 \oplus 3$. Now for up to one fault at any of the circled locations, an input error is corrected, and an internal error leaves an output error of weight $0$ or $1$.}
\label{f:smsintro1}
\end{figure}

\noindent
\textbf{Introduction.} The correction of errors in a quantum code requires a syndrome built from the measurement results of a sequence of stabilizers.
Since syndrome extraction is noisy, it is generally not sufficient to measure just a set of stabilizer generators, as shown in~\figref{f:smsintro1}.
Even one erroneous collected syndrome bit can result in an incorrect recovery, pushing the code into a state of logical error.  Instead, 
more stabilizers are redundantly measured to protect from quantum faults that cause syndrome bit flips.  The distance-$d$ surface code does this by measuring the stabilizer generators sequentially $\lceil \tfrac{d}{2} \rceil$ times, in a syndrome repetition code.  We may port this technique to the $k=2, 4$ and $6$ codes, but recent research has shown that these codes have very small stabilizer~\mbox{measurement sequences~\cite{delfosse2020short}.  }

\begin{figure*}
\subfigure[\label{f:smsintro} ]{\includegraphics[width=.27\textwidth]{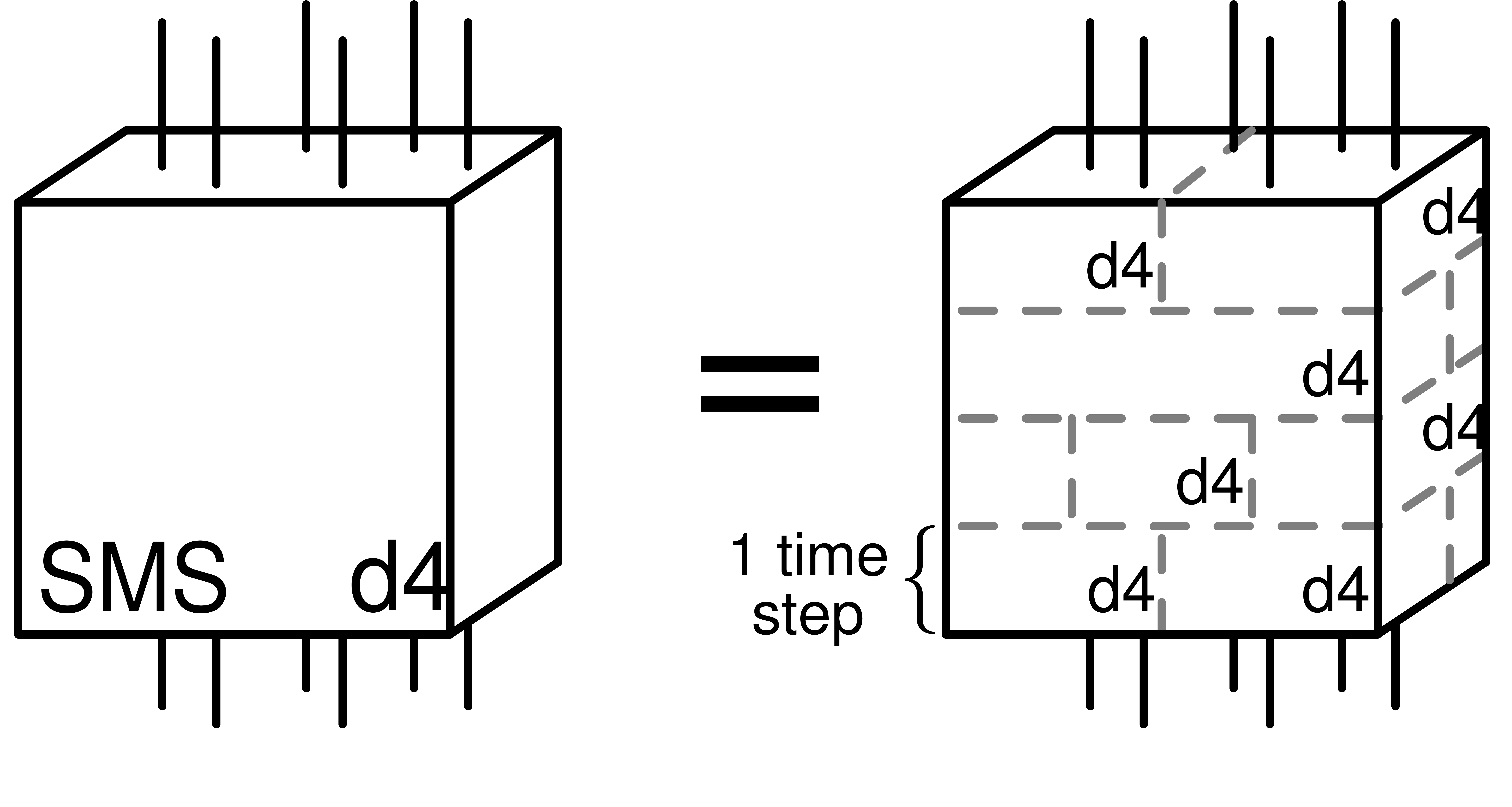}}
\hspace{8mm}
\subfigure[\label{f:smsrules} ]{\includegraphics[width=.67\textwidth]{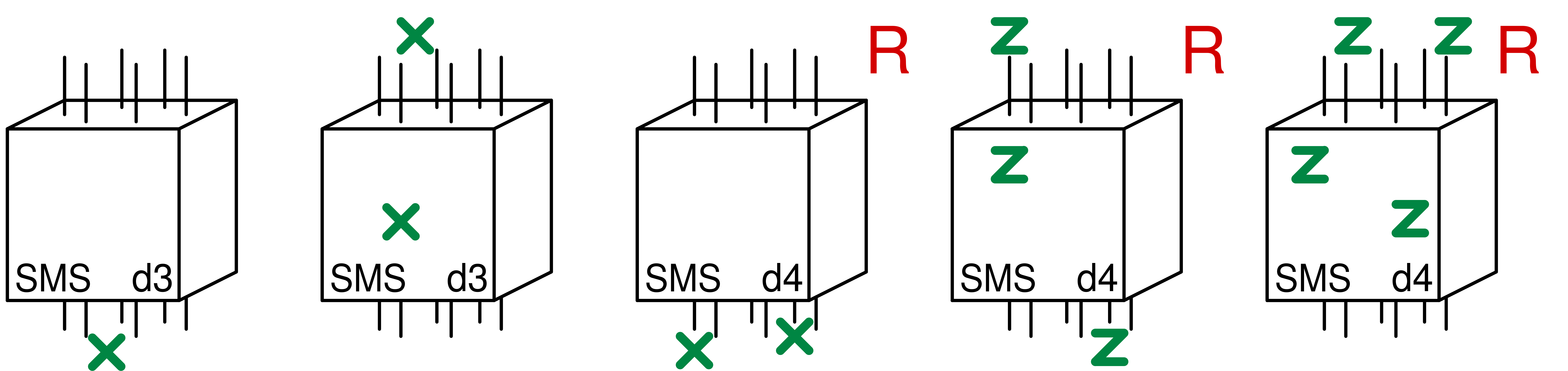}}
\caption{(a) A stabilizer measurement sequence (SMS) consists of multiple time steps of parallel stabilizer measurement circuits, where the end of a time step denotes the simultaneous measurement of all the ancilla qubits. (b) Rules for distance-$4$ fault tolerance---first two are sufficient for distance $3$: ($i$) An input $1$-qubit error must be corrected. ($ii$) $1$ internal fault must be corrected to an error of weight at most $1$. ($iii$) A $2$-qubit input error is rejected. ($iv$) $1$ input error and $1$ internal fault should be corrected to an error of weight at most $1$ or rejected. (v) $2$ internal faults must be rejected or propagate to an error of weight at most $2$.}
\vspace{-.3cm} 
\end{figure*}

The depth of a quantum circuit is generally calculated as the number of rounds of parallel two-qubit gates, since single-qubit gates are trivially short.  However, in current systems, the time needed for measurement dominates over the length of a CNOT
~\cite{Google2021, duke21BScode, honeywell21steane, riste20bitflipEC}.
Hence the focus shifts from minimizing CNOT depth to reducing the rounds of measurements needed for error correction.  We therefore denote by ``time step'' the time needed to measure a set of stabilizers in parallel, as shown in~\figref{f:smsintro}.
In addition to finding short fault-tolerant sequences of stabilizers, we carefully parallelize their measurement circuits to~\mbox{further speed up error correction.}

\pagebreak

\medskip
\noindent
\textbf{Fault tolerance rules.}
We follow the `exRec' formalism of Ref.~\cite{AliferisGottesmanPreskill05} to determine rules for fault-tolerant error correction, as shown in \figref{f:smsrules}.  For distance-three fault tolerance, only two rules are needed.  If the input to an error correction block has a weight-one error and there are no internal faults, the syndrome must be sufficient to correct back to the codespace. This is actually the basic rule for an ideal error correction block.  If there are no input errors and one internal fault occurs, the weight of the output error after recovery should be at most one.

For distance-four fault tolerance, we must consider the effect of up to two input errors or internal faults. If the input error has weight two and there are no internal faults, then the stabilizer measurement sequence must detect the error and restart the computation.  
If there is a weight-one input error and an internal fault, either the computation is restarted, or the output of the error correction block must have error of $X$ and $Z$ weight at most one. 
Finally, if two internal faults occur with no input error, either the computation is restarted, or the output error must have weight at most two. 

(In the last four rules, the resulting syndrome must never be equivalent to a weight-one input error on a different qubit. This ensures that every weight-one input error can be reliably corrected.)

\medskip
\noindent
\textbf{Solutions.} To perform distance-four fault-tolerant error correction on the layout of~\figref{f:layout}, we consider measuring stabilizers only of the form given in~\secref{s:SMC}. The goal is then to devise short parallel stabilizer measurement sequences occupying the fewest time steps while satisfying the fault tolerance rules above.  For each of the newly proposed codes, the sequences in~\figref{f:codes} were found using randomized search and subsequent minor alterations.

The $k = 2$ code measures ten $X$ (or $Z$) stabilizers over five time steps, hence recovery occurs every ten time steps. On the other hand, the $k = 6$ code contains no stabilizers of weight less than eight, so parallelism is difficult. Here, seven $X$ (or $Z$) stabilizers are measured over seven time steps, for a total $14$ time steps between recoveries. The distance-four surface code measures all its stabilizer generators in two time steps. The first is used to measure the nine weight-four stabilizer generators using the nine ancilla qubits, and the second time step is used to measure the boundary weight-two stabilizers. Recovery occurs after two fresh syndrome layers are measured, at a frequency of four time steps. 

\medskip
\noindent
\textbf{Rejection decoding with the surface code.}  Many algorithms exist to decode errors on the surface code.  With the hope of testing larger surface codes with rejection decoders, we implemented a version of the union-find decoding algorithm~\cite{delfosse2017almostlinear}.  To apply postselection rules for distance-four fault tolerance, we start at the end of union-find, with a set of edges representing $X$ or $Z$ faults. 
If there are three or more edges, we immediately reject and restart.  If there are two or fewer edges, some choices can be made.  First, if the two faults are well separated in time, the older fault can be reliably corrected.  We experimented with different separation lengths, and settled on a configuration that rejects least often, but has the highest $O(p^3)$ logical errors. Second, measurement faults do not contribute to data errors.  They are exempted from postselection by projecting edges onto the $2$D plane. Evidently, there is a lot of scope for further improvements to the postselection rules.  Research also needs to be done on how to generalize these rules to higher-\mbox{distance surface codes.}

\section{Results}\label{s:results}

\noindent
\textbf{Noise model.} For simulation, we consider independent circuit-level noise, as described below:

\begin{itemize}[leftmargin=*]
\item With probability $p$, the preparation of $\ket 0$ is replaced by $\ket 1$ and vice versa---similarly $\ket +$ and $\ket -$.
\item With probability $p$, $\pm X$ or $\pm Z$ measurement on any qubit has its outcome flipped.
\item With probability $p$, a one-qubit gate is followed by a random Pauli error drawn uniformly from $\{ X, Y, Z\}$.
\item With probability $p$, the two-qubit CNOT gate is followed by a random two-qubit Pauli error drawn uniformly from $\{ I, X, Y, Z\}^{\otimes 2} \setminus \{I \otimes I \}$.
\item After each time step, with probability $p(1+m/10)$, each data qubit is acted upon by a random one-qubit Pauli error drawn uniformly from $\{X, Y, Z\}$.  (A time step denotes one round of parallel stabilizer measurements of maximum CNOT depth $m$, as in~\secref{s:SMS}.)
\end{itemize}

The rest error rate models the observed performance of current-day quantum systems, where the time taken to measure an ancilla qubit is long compared to the CNOT gate time.  We model the rest error rate during measurement as $p$, and during CNOT gates as $p/10$.  Even with dynamical decoupling~\cite{ViolaKnillLloyd99DD}, the error incurred by the idle data qubits can be quite high.  

\medskip
\noindent
\textbf{Normalized logical error rate.} 
The logical error rate of fault-tolerant storage can be estimated by checking for a logical error after each block of error correction.
However, different codes correct errors at different frequencies---once every four time steps for the surface code, but fourteen for the $k=6$ code. 
To compare the codes on a similar time scale, we normalize the logical error rates~\mbox{with respect to time step.}

\begin{figure}
\includegraphics[width =0.46\textwidth]{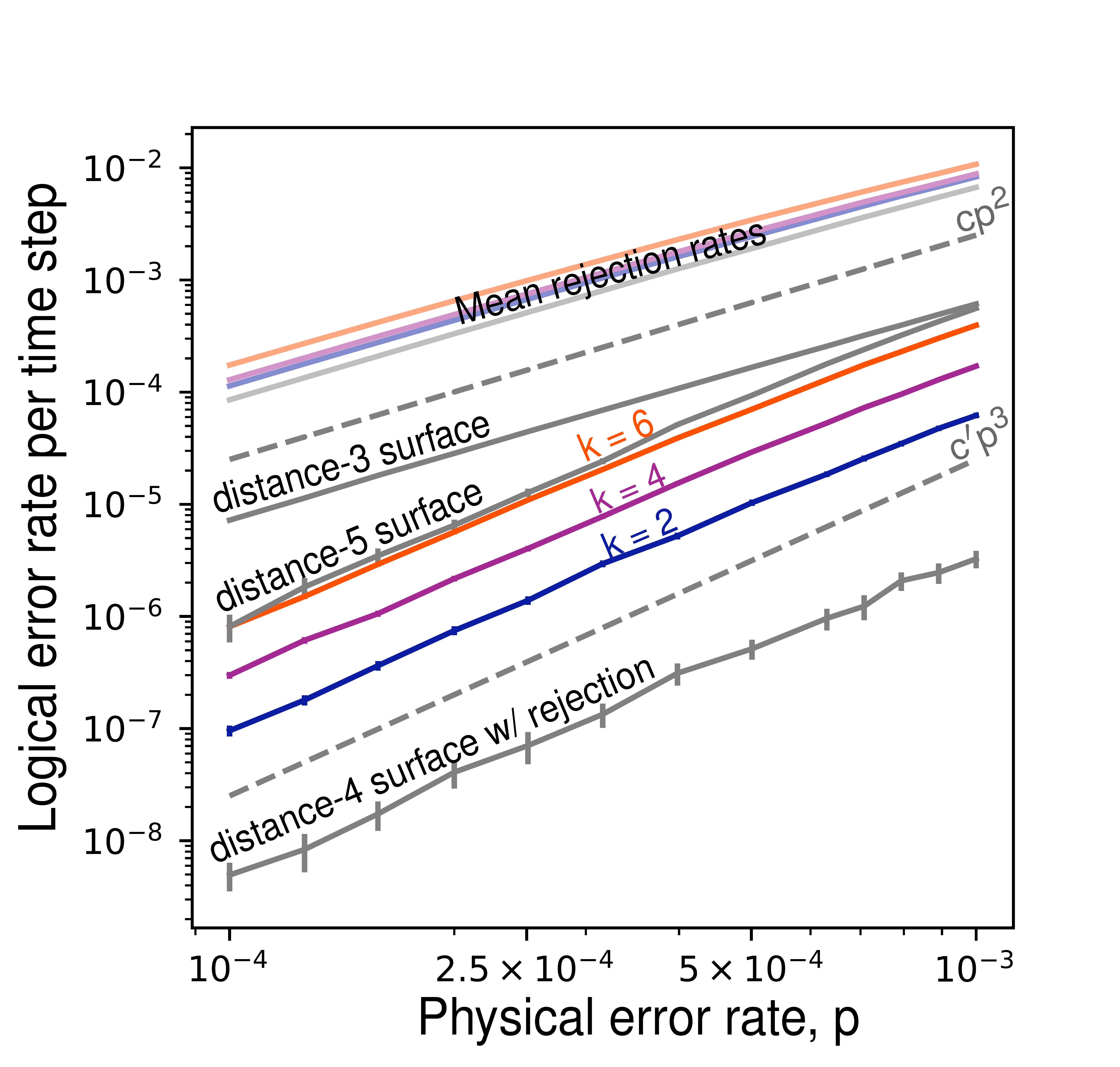}
\caption{$O(p^3)$ scaling of $X$ logical error rate and $O(p^2)$ scaling of rejection rate, with error bars, for the distance-four codes. The distance-three and distance-five surface codes are shown for comparison. The new codes have logical error rate per time step as low as $1/10$th the distance-five surface code.
The distance-four surface code is as low as $1/100$.}
\label{f:scaling}
\vspace{-0.3cm} 
\end{figure}

We plot the logical error rate per time step in~\figref{f:scaling}, where we show that a distance-four surface code has a storage error rate of $O(10^{-9})$, for a CNOT gate error rate of just $10^{-4}$.  Even with the infidelity of current day CNOTs, $\sim 10^{-3}$, we show logical error rates approaching $10^{-6}$.  These results demonstrate the benefits of postselection.

\medskip
\noindent
\textbf{Cumulative logical error probability.} 
The mean rejection rates in~\figref{f:scaling} provide a good comparison of how often the different codes reject, but do not accurately describe behavior for bounded-length computation.  Here, a more useful metric is the probability of acceptance, $P_a(t)$, which is how often a $t$-time-step computation completes.  This quantity can be estimated empirically by simulating the application of noisy error correction to an initial state for bounded time, which we denote as a simulation `run'.  If $R$ is the total number of executed runs and $R_{a}(t)$ is the number of runs that have not rejected until time step $t$, 
\begin{equation}
P_a(t) = \frac{R_{a}(t)}{R}  \, .
\end{equation}

Similar to the rejection rate, the logical error rate per time step is indicative of the frequency of logical errors, but does not help to understand the drawbacks of postselection. We again refer to a cumulative metric, the probability of a logical error after $t$ time steps of error correction, empirically given by
\begin{equation}
P_{L}(t)= \frac{R_{L}(t)}{R}  \, ,
\end{equation}
where $R_{L}(t)$ is the number of runs in a state of logical error at time step $t$.  For even distance codes, one must instead look at the probability of logical error conditioned on acceptance, which is calculated as
\begin{equation}
P_{L | a}(t) =  \frac{P_{L}(t)}{P_{a}(t)}  = \frac{R_{L}(t)}{R_{a}(t)}  =  \frac{R_{L}(t)}{R \, P_{a}(t)} \, .
\end{equation}
For even distance codes with postselection, $P_a(t) < 1$ and so $P_{L | a}(t) > P_{L}(t)$.  For odd distance codes that do not reject, $P_{L | a}(t) = P_{L}(t)$.

The above formula holds only for a single code patch.  The probability of logical error while using multiple patches can be upper bounded from the data for a single patch as 
\begin{equation}
P_{L | a}(t,c) \leq \frac{c \, R_{L}(t)} {R \, P_a^c(t)} \, ,
\end{equation}
where $c$ is the number of code patches used. Note that the number of logically incorrect runs grows linearly with the number of patches, but the probability of acceptance of multiple patches is the probability that every patch has accepted.

\begin{figure*}
\vspace{-0.5cm}
\includegraphics[width=\textwidth]{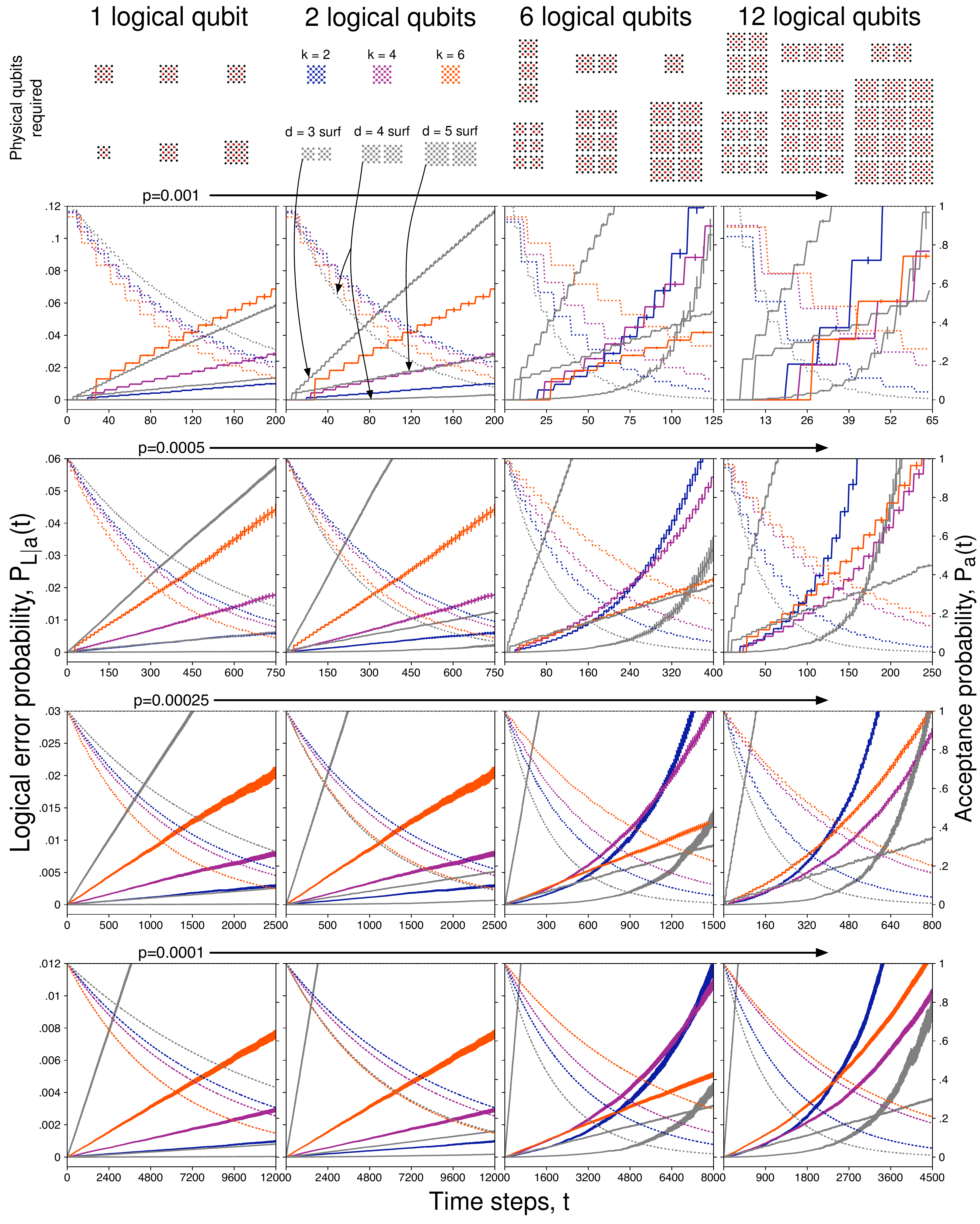}
\caption{Probability of logical error (solid) and acceptance (dotted) for $t$ time steps of error correction on six codes, as a function of physical error rate (row) and desired logical qubits (column).  
The three colored curves correspond to the ${\color{curblue} k = 2}$, ${\color{curpurple} k = 4}$ and ${\color{curorange} k = 6}$ codes and the three gray curves are the surface codes.  The graphs (especially for few time steps) look like a step function because the code patches are checked for logical errors only after blocks of error correction, not time steps.  The top row compares the number of physical qubits required to achieve the desired number of logical qubits.}
\label{f:PLEt}
\end{figure*}

\medskip
\noindent
\textbf{Discussion} 
We simulated fault-tolerant error correction of the codes in~\figref{f:codes} for up to 12000 time steps at error rate $p \in \{ 0.001, 0.0005, 0.00025, 0.0001\}$. Using the empirical formulae above, we then plotted in~\figref{f:PLEt} the probability of logical error conditioned on acceptance and the probability of acceptance for one, two, six or twelve logical qubits.  Note that some plots look discontinuous. This is because we only check for logical errors and reject at the end of an error correction block.  Above the graphs of~\figref{f:PLEt}, we compare the number of physical qubits required for each code.

\begin{table}
\caption{\label{f:k1p0_001}
Error correction for $1$ logical qubit at $p = 0.001$. The probability of logical error and acceptance are shown for $80$ and $200$ time steps.  Each code uses one patch of qubits.  The distance-four surface code has the lowest logical error probability for short computations.  
}
\vspace{-0.9mm}
\hspace{-2.1mm}
\includegraphics[width=0.49\textwidth]{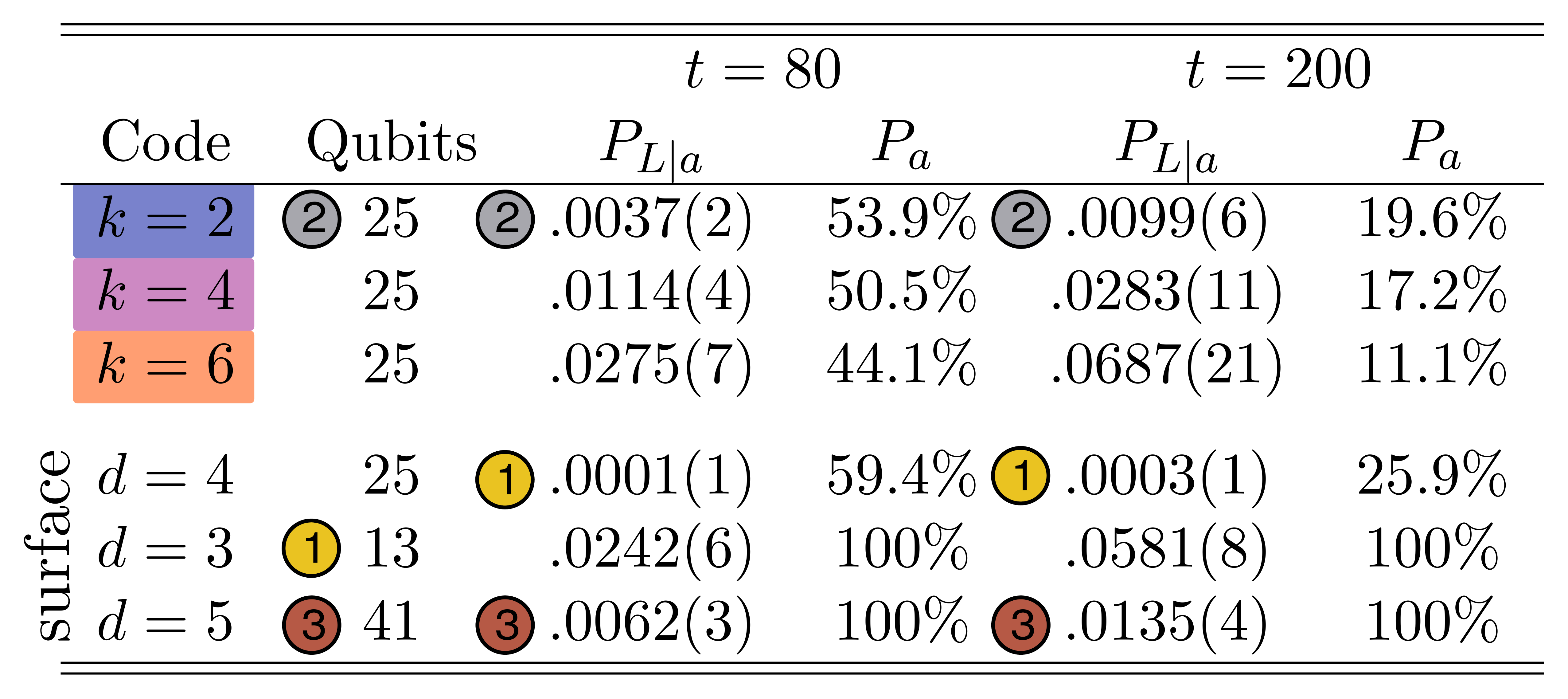}
\vspace{-0.5cm} 
\end{table}

There are many things to be learned from~\figref{f:PLEt}. To start, the first column of graphs shows how a single patch of each code fares against the others, for different error rates.  The $d = 4$ surface code with rejection boasts the lowest logical error probability overall and has the highest acceptance rates among all the even-distance codes.  The logical error probability of the $k = 2$ code actually matches the distance-five surface code, even though it encodes twice as much information.  This is also apparent from~\tabref{f:k1p0_001}, where we show the probability of acceptance and logical error for one logical qubit at $p = 10^{-3}$.

\begin{table}
\caption{\label{f:k2p0_0005}
Error correction for $2$ logical qubits at $p = 0.0005$.  The probability of logical error and acceptance are shown~for $300$ and $750$ time steps.  The surface codes require more than one patch of physical qubits.  Among the new codes, the $k = 2$ color code has few large stabilizers and a fast sequence. These advantages help it achieve the lowest logical error probability at the highest acceptance rates.}
\vspace{-0.9mm}
\hspace{-2.1mm}
\includegraphics[width=0.49\textwidth]{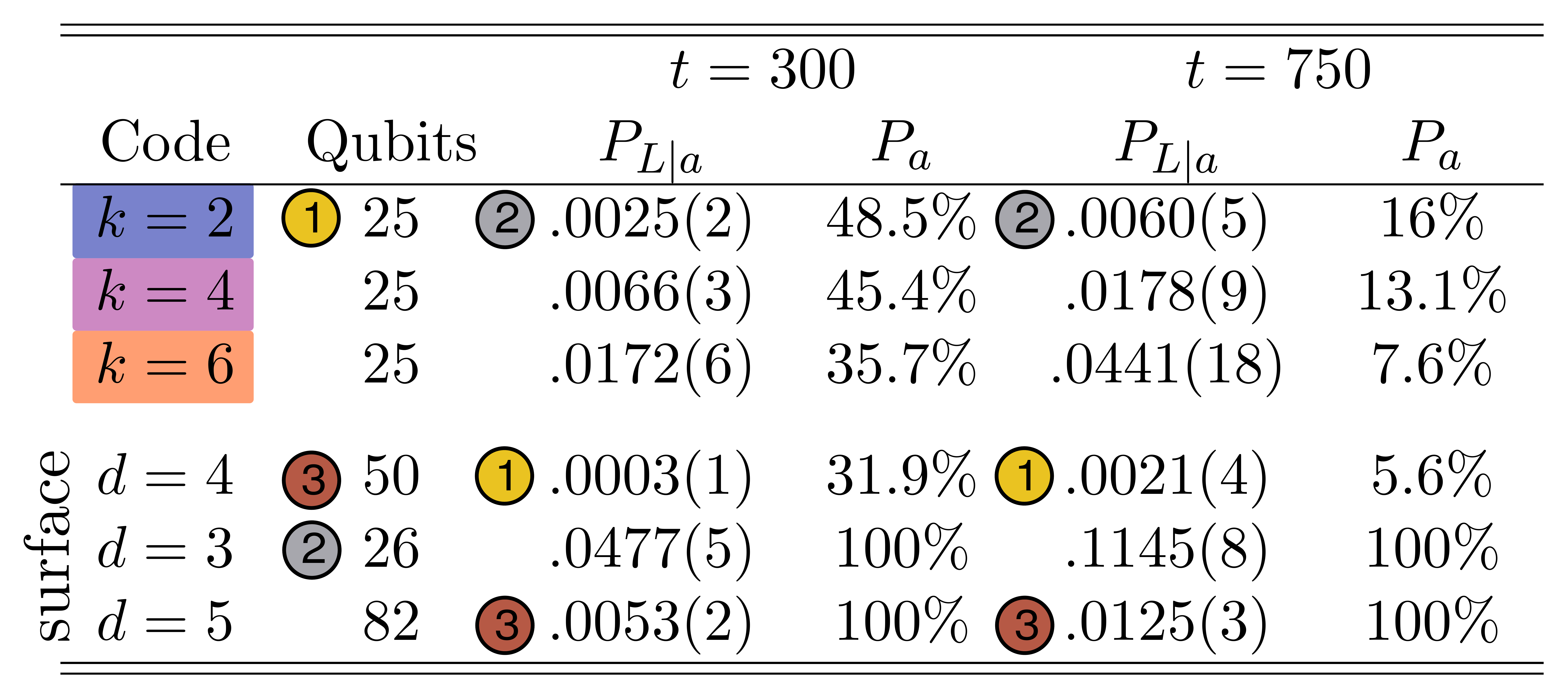}
\vspace{-0.8cm}  
\end{table}

For two logical qubits, (second column of graphs), the surface codes need two patches of qubits, hence the probability of logical error doubles and the acceptance is squared. The distance-four surface code now has the lowest acceptance probability among the distance-four codes.  We keep the range of time steps consistent between the first and second columns to show that the curves for the multi-qubit codes are unchanged.  As shown in~\tabref{f:k2p0_0005}, for two logical qubits, the $k = 2$ code halves the logical error probability of the $d = 5$ surface code, using fewer than one-third as many physical qubits. 

Going further, we have analyzed error correction for six and twelve encoded qubits, as shown in the last two columns.
With only one-tenth of the physical overhead, a single $k = 6$ code patch rivals the performance of six patches of the $d = 5$ surface code.  The single patch of the $k=6$ code even outperforms the $k = 2$ and $k = 4$ codes, but this is precisely because only one code patch is used. When multiple code patches are used for many logical qubits, the acceptance rate of the distance-four codes drops exponentially.  This is also observed in~\tabref{f:k6p0_00025} and~\tabref{f:k12p0_0001}, as the acceptance probability of the distance-four surface code quickly approaches zero.

In the last column of graphs, we compare statistics for twelve logical qubits. Although current NISQ systems only protect one logical qubit, our results show that just $50$--$75$ physical qubits are sufficient for twelve logical qubits. 
In this regime, the $k = 4$ code achieves lower logical error probability than the $k=6$ code with only $50\%$ more overhead. Unfortunately at longer time scales, postselection sharply increases the logical error probability, rendering the distance-four codes much less useful.

All simulations in this paper, developed in Python, were executed on the USC Center for Advanced Research Computing (CARC) high-performance computing cluster. The simulations used over one million minutes of CPU core time on Intel Xeon processors operating at $2.4$ GHz.

\medskip
\noindent
\textbf{Take-home message.} Postselection can play a crucial role in reducing logical error rates.  
However, when logical information is stored for too long, it is likely to be wiped and reset. 
This is okay for some algorithms: applications with low depth, like variational algorithms~\cite{cerezo2020variational, bharti2021noisy}, or those that are designed with rejection, like magic state distillation~\cite{BravyiKitaev04magic}.
If only one or two qubits are required, the distance-four surface code and the $k=2$ code offer very low probability of logical error. For more qubits, we advise using the $k=4$ or $k=6$ codes, as they use far fewer physical resources to achieve competitively low logical error.  
We show that $50$--$75$ good physical qubits are sufficient to correct errors on twelve logical qubits. Even at a CNOT error rate as high as $5\times 10^{-4}$, error correction up to $100$ time steps can be run with error probability as low as $1\%$.

\begin{table}
\caption{\label{f:k6p0_00025}
Error correction for $6$ logical qubits  at $p = 0.00025$.  The probability of logical error and acceptance are shown~for $700$ and $1500$ time steps.  The $k = 6$ code requires one-tenth the physical qubits as the distance-$5$ surface code, while nearly matching the logical error probability.}
\vspace{-0.9mm}
\hspace{-2.1mm}
\includegraphics[width=0.49\textwidth]{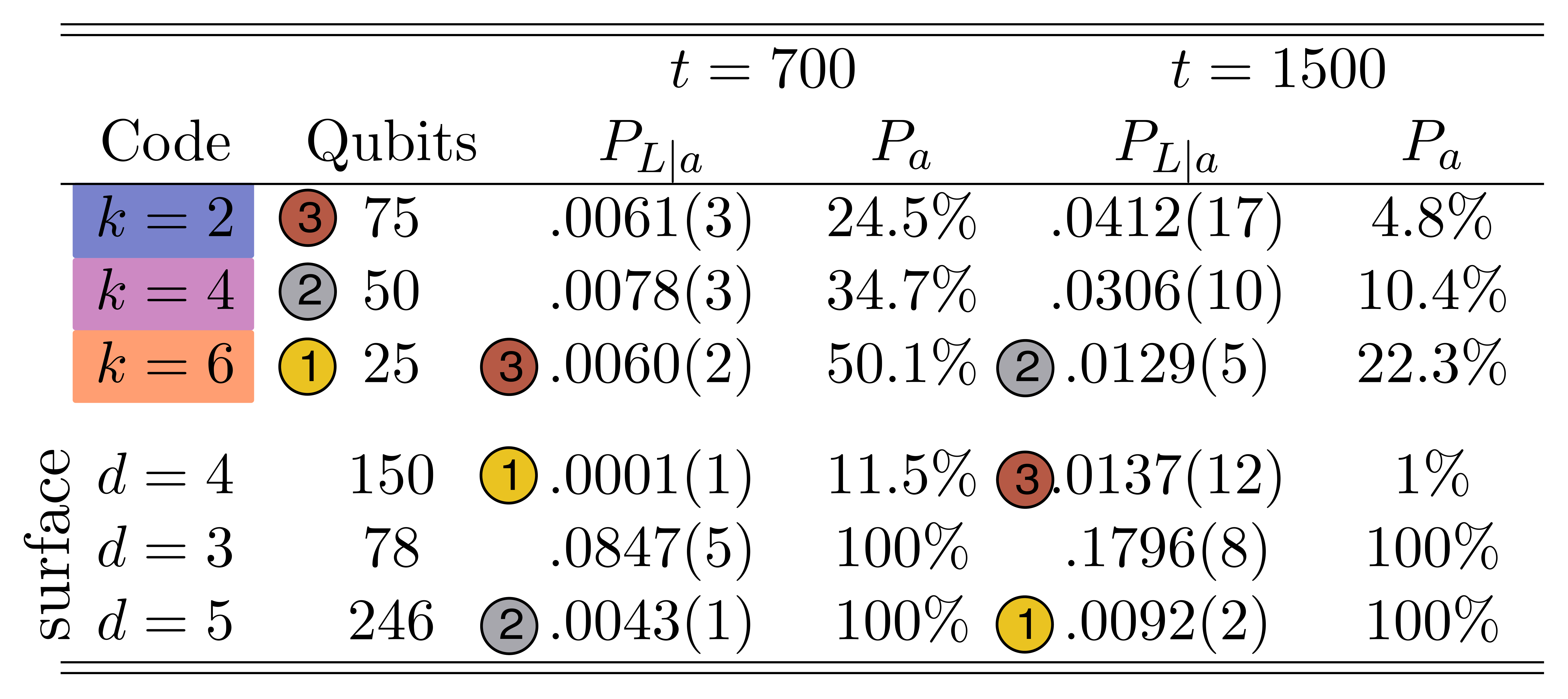}
\vspace{-0.6cm} 
\end{table}

\begin{table}
\caption{\label{f:k12p0_0001}Error correction for $12$ logical qubits at $p=0.0001$.  The probability of logical error and acceptance are shown~for $1800$ and $4500$ time steps.  The $k=4$ code is well-balanced, achieving competitive logical error rates with low~\mbox{qubit overhead.}}
\vspace{-0.9mm}
\hspace{-2.1mm}
\includegraphics[width=0.49\textwidth]{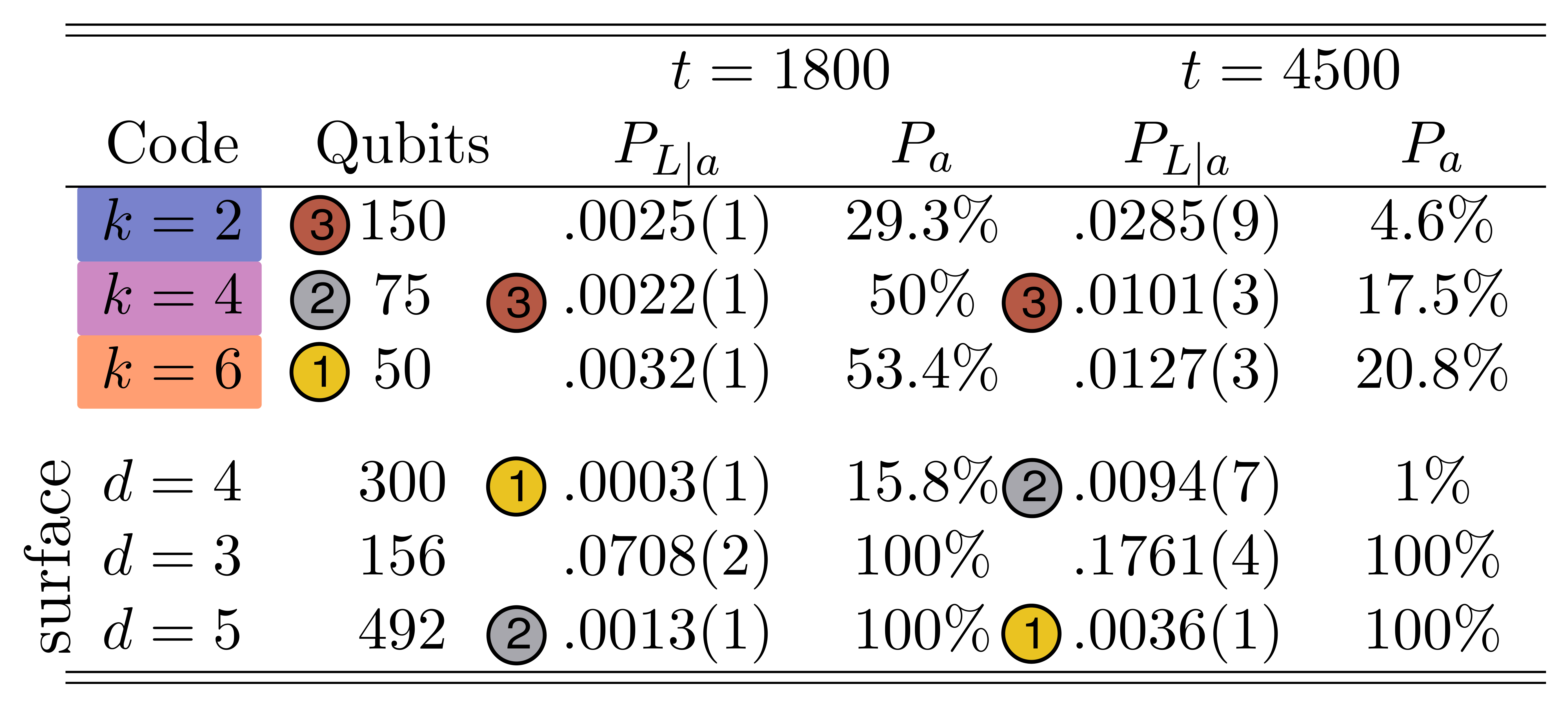}
\end{table}

\section{Future work}
\label{s:future}

\begin{figure}
\includegraphics[width=.38\textwidth]{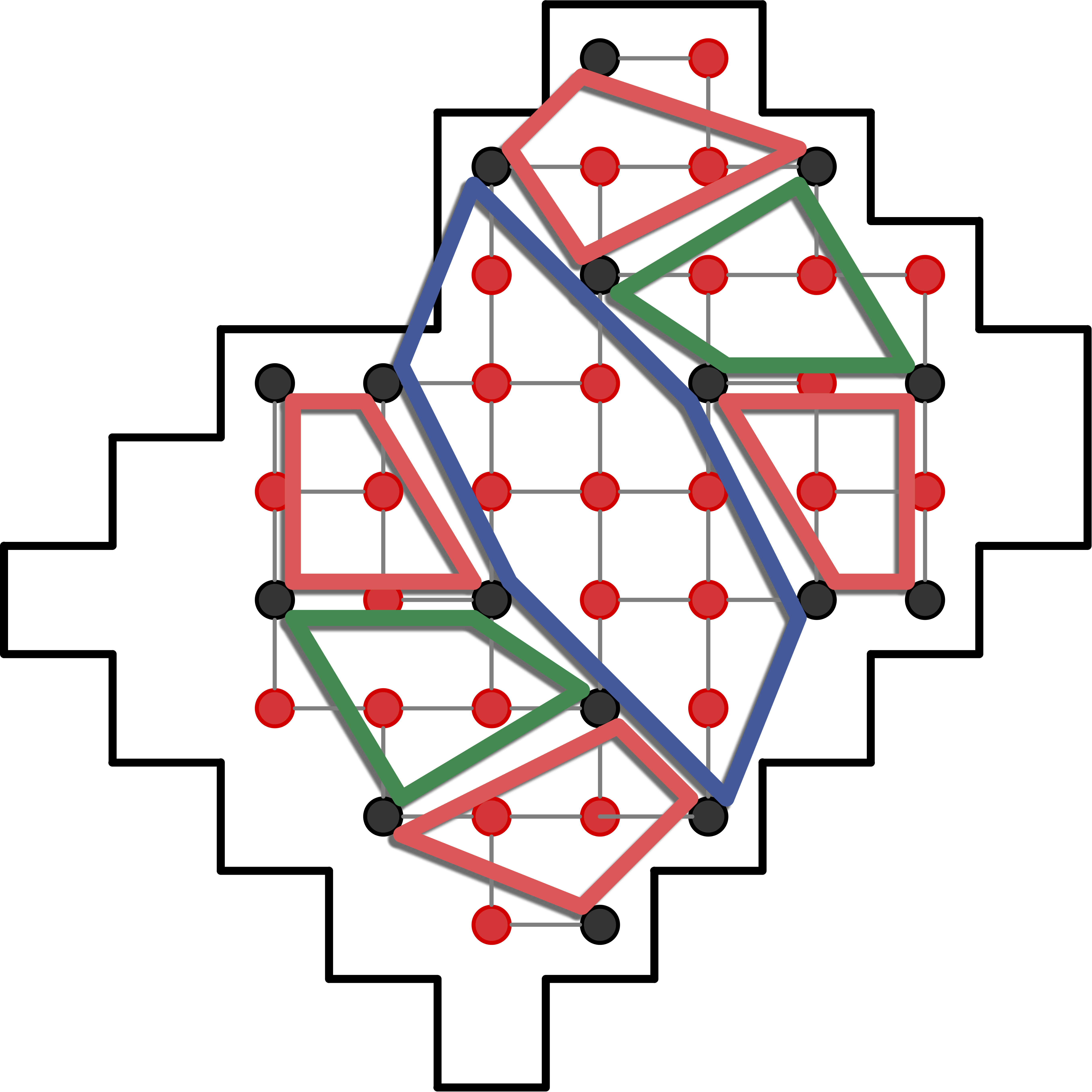}
\caption{A degree-$4$ layout for flag-fault-tolerant error correction of the $k = 2$ code, using $43$ of the $53$ qubits on the Google Sycamore lattice. The stabilizer generators of the code are overlaid. Note that qubits have degree $4$ only in the ancillas measuring the weight-$8$ stabilizer, but elsewhere the maximum qubit degree is $3$. It may be possible for the $k=2$ subsystem code with only weight-$4$ stabilizers and gauges to fit on a layout of maximum degree $3$.} 
\label{f:degree4k2}
\end{figure}

In this paper, we show how to perform fault-tolerant storage with $16$-qubit codes. There are two immediate roadblocks en route to universal fault-tolerant quantum computation. Currently, no devices exist with the layout of~\figref{f:layout}, so until they are fabricated, we turn to other layout improvements.
The middle ancilla qubit in~\figref{f:layout} is connected to eight neighboring qubits.  
However, careful analysis and modification of the stabilizer measurement routines may yield solutions that only require maximum qubit degree five or six.
This may not be interesting for densely-connected ion trap quantum computers, but is necessary in superconducting architectures to maintain low cross-talk.  Alternatively, if we are allowed extra ancilla qubits, we show that maximum degree four is possible, as in the Google Sycamore lattice of~\figref{f:degree4k2}.  The stabilizer measurement circuits are all fault-tolerant to distance four, but since all the stabilizer generators are measured simultaneously, error decoding will require new strategies.  The weight-four stabilizers can be measured using the circuit in~\figref{f:wt4circ}, but the weight-eight stabilizer requires a new circuit, as we detail in~\appref{s:wt8corrs}.  In~\figref{f:degree4k2}, the only qubits with degree-four connectivity are the ancillas used for measuring the weight-eight stabilizer. For systems with high crosstalk, qubits of degree three may be sufficient to correct errors on the $k=2$ subsystem code, since errors can be corrected by measuring only weight-four~operators.

On the theoretical front, we must develop encoding circuits and a universal logical gate set.  States may be prepared by either using flags for fault-tolerance, or by combining patches of distance-two code states into a distance-four state.  For fault-tolerant universal computation, one possible route is teleportation and logical measurements with distilled magic states.  In fact, logical measurements can be performed along with error correction~\cite{delfosse2020short}.  Another route to universality is to use transversal multi-qubit gates between vertically stacked code patches.  It may be possible for gates like the CCZ to induce magic~\cite{PaetznickReichardt13universal}, as the required $\llbracket 15,7,3 \rrbracket$ code can be obtained by puncturing the $\llbracket 16,6,4 \rrbracket$ code. If the error introduced by logical operations is kept low, many logical gates can be applied every time step, allowing high-depth logical circuits.

Near the threshold of the odd distance surface codes, postselection on the distance-two and -four  variants shows reduced logical error rates.  With higher distance, the compounding effect is larger, meaning distance-eight or -ten surface codes may be sufficient for very precise computations.  At higher distance, rejections also become exceedingly rare, increasing the possible duration of computations.  
For larger patches on lattices like~\figref{f:layout}, more qubits can be encoded at high distance.

The biggest difficulty will then be in performing operations on or between different logical qubits in the same patch. 
Another avenue to pursue is concatenation. This technique can combine the low logical error rates of the surface code with the high encoding rates of block codes.  

\medskip
\medskip
\section{Conclusion}

We have simulated postselected fault-tolerant error correction for distance-four $16$-qubit codes. Qubits are positioned on a $2$D grid, and the only interactions allowed are local CNOT gates. Flag qubits are used to measure high-weight stabilizers, allowing for small error correction circuits for block codes encoding up to six logical qubits. 
With a phenomenological noise model, it is not possible to correct errors reliably using a syndrome of just the stabilizer generators.  Instead, we have proposed longer sequences of stabilizers that use redundant measurements to detect syndrome bit flips.  These new techniques are still in early development, and we have suggested multiple improvements for each of them along the way.

We have shown a variety of results interpolating between low logical error rates and low physical overhead. The downside of using postselection is that the logical qubits cannot be stored for too long, but low-depth NISQ algorithms are definitely possible.  For these shorter time scales, we have compared the logical error probability to a distance-five surface code: \textit{i)} The distance-four surface code has at least a twenty-five-fold decrease in logical error rate, \textit{ii)} A $k=2$ color code with weight-eight stabilizers halves the probability of a logical error with only $30\%$ of the qubits, and \textit{iii)} A $k=6$ block code matches performance with just $25$ qubits as opposed to $246$. Using just $25$--$75$ physical qubits, we have demonstrated that it is possible to protect well six to twelve logical qubits.

\section{Acknowledgments}

The authors thank Todd Brun, Anirudh Lanka, Dripto Debroy, Nicolas Delfosse and Zhang Jiang for useful conversations.  Research supported by Google and by MURI Grant FA9550-18-1-0161.  This material is based on work supported by the U.S. Department of Energy, Office of Science, National Quantum Information Science Research Centers, Quantum Systems~Accelerator.  The authors acknowledge the Center for Advanced Research Computing (CARC) at the University of Southern California for providing computing resources that have contributed to the research results reported within this publication. URL: https://carc.usc.edu.

\ifx\compilefullpaper\undefined  
\bibliographystyle{halpha-abbrv}
\bibliography{q}

\appendix

\vspace{-0.6cm}
\section{Corrections and rejections for weight-eight stabilizer measurements}
\label{s:wt8corrs}

\vspace{-0.1cm}

\begin{figure}
\hspace{-5mm}
\subfigure[\label{f:degree4k2wt8circ} ]{\includegraphics[width=.375\textwidth]{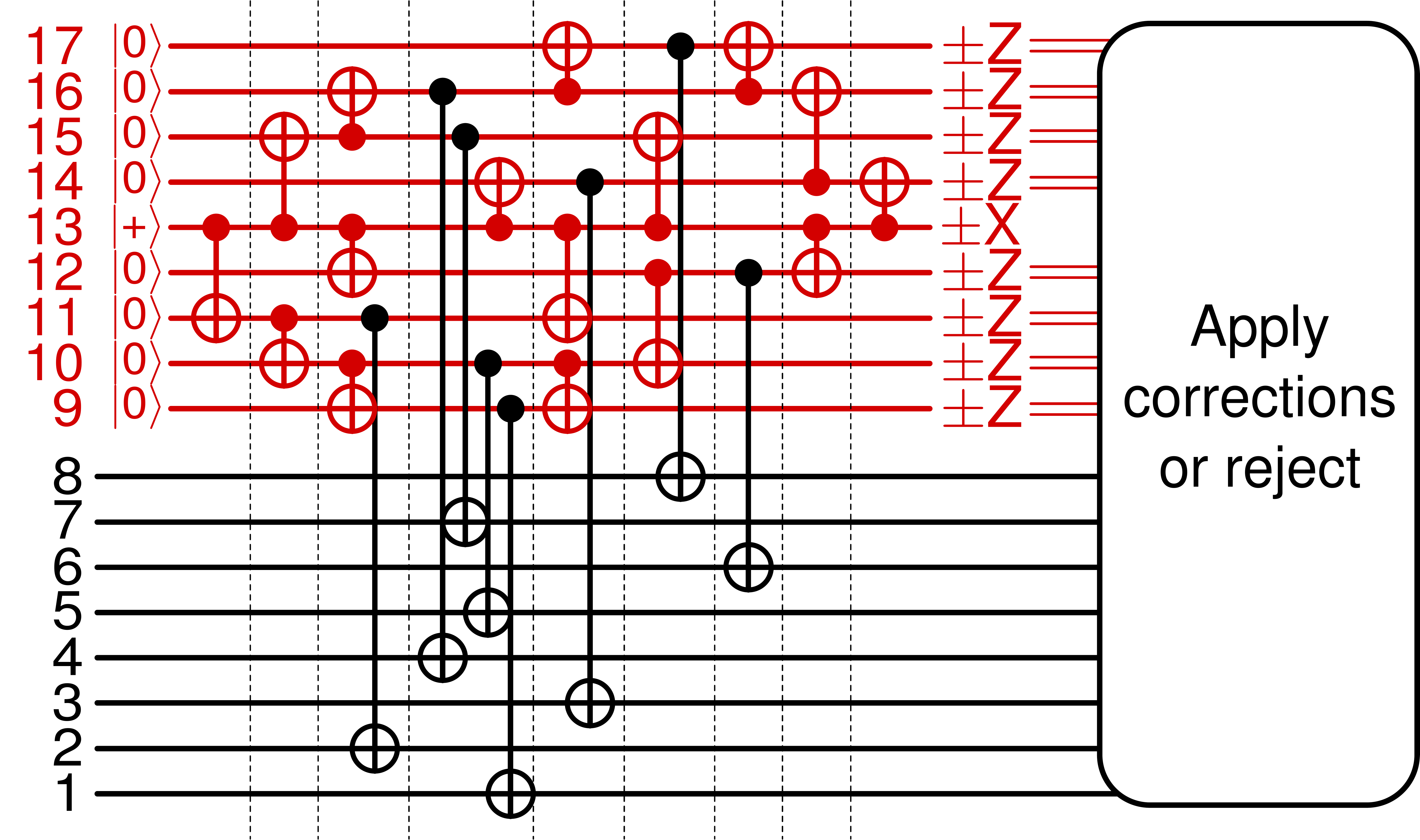}}
\subfigure[\label{f:degree4k2wt8layout} ]{\includegraphics[width=.118\textwidth]{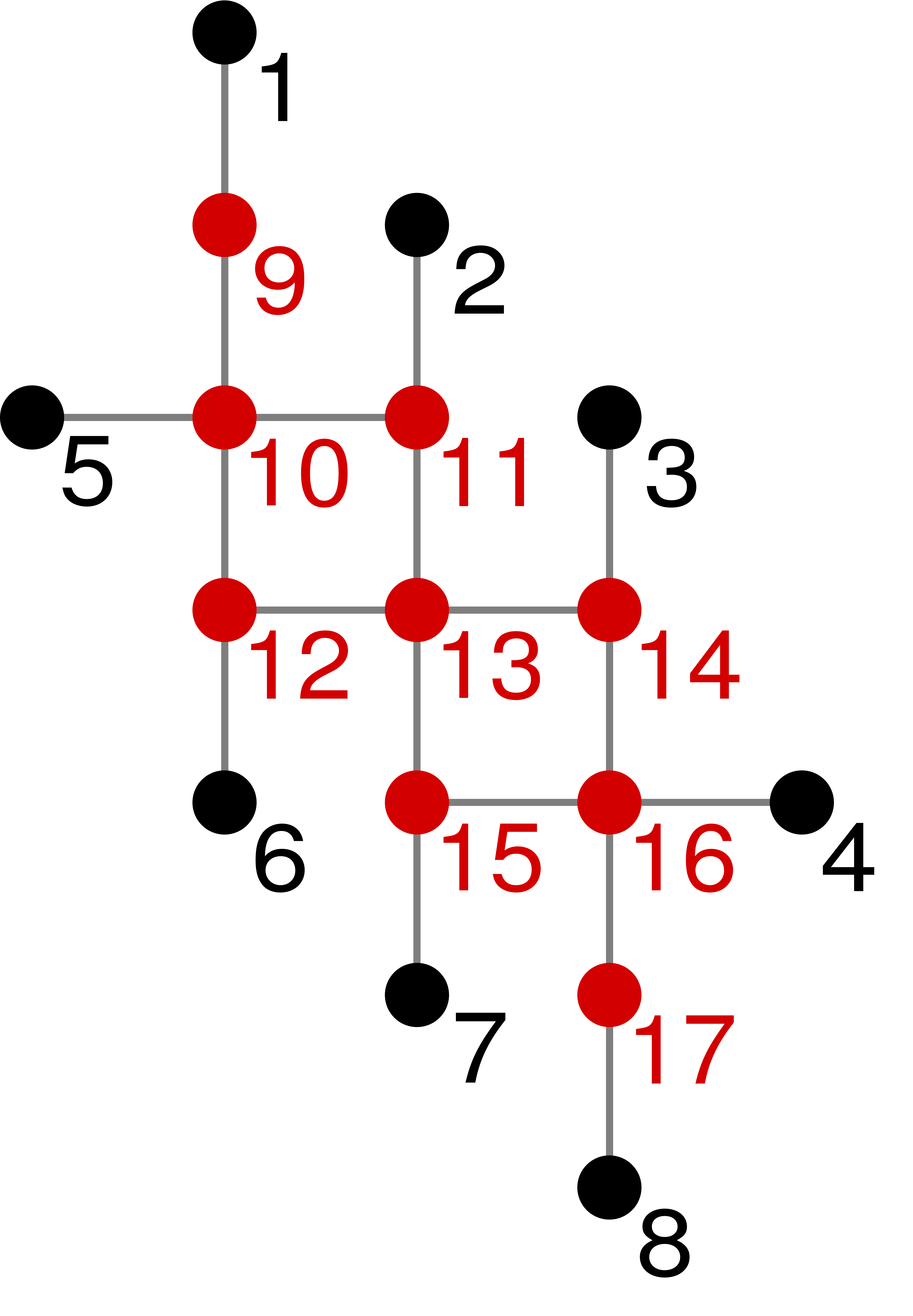}}
\caption{(a)~Distance-4 fault-tolerant circuit for measuring a weight-8 stabilizer on a square lattice layout, as arranged in~(b).}
\label{f:degree4k2wt8}
\end{figure}

The corrections and rejections for the fault-tolerant weight-eight stabilizer measurement circuit in~\figref{f:wt8circ} are shown below.

\noindent
\setlength{\tabcolsep}{5pt}
\begin{tabular}{cc|cc}
Raised flags & Correction & Raised flags & Correction  \\
\hline 
\{$13$\} & \{$6$\} & \{$12,13$\} & \{$6$\} \\
\{$9,12$\} & \{$1,5$\} & \{$12,13,14$\} & \{$6,7,8$\} \\
\{$9,11,12,14$\} & \{$1,4,5$\} & $$ & $$ \\ 
\end{tabular} 

\vspace{0.1cm}

\noindent
\begin{tabular}{c}
Rejections \\
\hline
\multicolumn{1}{p{0.97\linewidth}}{\{$9,11$\}, \{$9,13$\}, \{$9,14$\}, \{$11,12$\}, \{$12,14$\}, \{$13,14$\}, \{$9,11,12$\}, \{$9,11,14$\}, \{$9,12,13$\}, \{$9,12,14$\}, \{$9,13,14$\}, \{$11,12,13$\}, \{$11,12,14$\}, \{$11,13,14$\}, \{$9,11,12,13$\}, \{$9,11,13,14$\}, \{$9,12,13,14$\},  \{$11,12,13,14$\}, \{$9,11,12,13,14$\} }
\end{tabular}

\vspace{0.25cm} 

For the layout described in~\figref{f:degree4k2}, the weight-eight stabilizer is measured fault-tolerantly with the circuit in~\figref{f:degree4k2wt8}, with associated corrections and rejections tabulated below.

\vspace{0.25cm} 

\noindent
\setlength{\tabcolsep}{5pt}
\begin{tabular}{cc|cc}
Raised flags & Correction & Raised flags & Correction  \\
\hline 
\{$10$\} & \{$1$\} & \{$11$\} & \{$2$\} \\
\{$12$\} & \{$6$\} & \{$14$\} & \{$8$\} \\
\{$15$\} & \{$7$\} & \{$16$\} & \{$8$\} \\
\{$10,11$\} & \{$1,2$\} & \{$10,12$\} & \{$1,6$\} \\
\{$15,16$\} & \{$7,8$\} & \{$10,11,15,16$\} & \{$3$\}  \\ 
\end{tabular} 

\noindent
\begin{tabular}{c}
Rejections \\
\hline
\multicolumn{1}{p{0.97\linewidth}}{
 \{$9, 11$\},  \{$ 9, 12$\},  \{$9, 15$\},  \{$9, 16$\},  \{$9, 17$\},  \{$10, 14$\},  \{$10, 15$\},  \{$10,  
16$\},  \{$10, 17$\},  \{$11, 12$\},  \{$11, 15$\},  \{$11, 16$\},  \{$11, 17$\},  \{$12, 15$\},  \{$12,  
16$\},  \{$12, 17$\},  \{$14, 15$\},  \{$15, 17$\},  \{$9, 10, 11$\},  \{$9, 10, 12$\},  \{$9, 10,  
15$\},  \{$9, 10, 16$\},  \{$9, 10, 17$\},  \{$9, 14, 16$\},  \{$9, 15, 16$\},  \{$10, 11,  
12$\},  \{$10, 11, 14$\},  \{$10, 11, 15$\},  \{$10, 11, 16$\},  \{$10, 11, 17$\},  \{$10, 12,  
14$\},  \{$10, 12, 15$\},  \{$10, 12, 16$\},  \{$10, 12, 17$\},  \{$10, 14, 16$\},  \{$10, 15,  
16$\},  \{$10, 16, 17$\},  \{$11, 12, 15$\},  \{$11, 12, 16$\},  \{$11, 14, 16$\},  \{$11, 15,  
16$\},  \{$12, 14, 16$\},  \{$12, 15, 16$\},  \{$14, 15, 16$\},  \{$14, 16, 17$\},  \{$15, 16,  
17$\},  \{$9, 10, 14, 16$\},  \{$9, 10, 15, 16$\},  \{$9, 11, 15, 16$\},  \{$10, 11, 12,  
14$\},  \{$10, 11, 14, 15$\},  \{$10, 11, 14, 16$\},  \{$10, 11, 15, 17$\},  \{$10, 11,  
16, 17$\},  \{$10, 12, 14, 15$\},  \{$10, 12, 14, 16$\},  \{$10, 12, 15, 16$\},  \{$11,  
12, 14, 16$\},  \{$12, 14, 15, 16$\},  \{$9, 10, 11, 15, 16$\},  \{$9, 11, 12, 15,  
16$\},  \{$10, 11, 12, 14, 15$\},  \{$10, 11, 12, 14, 16$\},  \{$10, 11, 12, 15,  
16$\},  \{$10, 11, 14, 15, 16$\},  \{$10, 11, 15, 16, 17$\},  \{$11, 12, 14, 15,  
16$\},  \{$11, 12, 15, 16, 17$\},  \{$9, 10, 11, 12, 15, 16$\},  \{$10, 11, 12, 14,  
15, 16$\} }
\end{tabular}

\hbox{}

\end{document}
\fi